\documentclass[preprint]{aastex6}
\usepackage{color}
\usepackage{amsmath}
\fullcollaborationName{The Friends of AASTeX Collaboration}

\begin{document}


\title{X-ray Studies of the Extended TeV Gamma-Ray Source VER~J2019+368}


\author{
T. Mizuno\altaffilmark{1}, 
N. Tanaka\altaffilmark{2}, 
H. Takahashi\altaffilmark{2},
J. Katsuta\altaffilmark{2},
K. Hayashi\altaffilmark{3}, and
R. Yamazaki\altaffilmark{4}
}

\email{mizuno@astro.hiroshima-u.ac.jp}


\altaffiltext{1}{Hiroshima Astrophysical Science Center, Hiroshima University, 1-3-1 Kagamiyama, Higashi-Hiroshima, Hiroshima, 739-8526, Japan}
\altaffiltext{2}{Department of Physical Science, Hiroshima University, 1-3-1 Kagamiyama, Higashi-Hiroshima, Hiroshima, 739-8526, Japan}
\altaffiltext{3}{Department of Physics, Nagoya University, Chikusa-ku, Nagoya 464-8602, Aichi, Japan}
\altaffiltext{4}{Department of Physics and Mathematics, Aoyama Gakuin University, 5-10-1 Fuchinobe, Sagamihara, 252-5258, Kanagawa, Japan}

\begin{abstract}

This article reports the results of X-ray studies of the extended TeV $\gamma$-ray source VER~J2019+368.
\textit{Suzaku} observations conducted to examine properties of the X-ray 
pulsar wind nebula (PWN) around PSR~J2021+3651
revealed that the western region of the X-ray PWN has a source extent of
$15' \times 10'$ with the major axis oriented to that of the TeV emission. 
The PWN-west spectrum was closely fitted by a power-law for absorption at
$N({\rm H}) = (8.2^{+1.3}_{-1.1}) \times 10^{21}~{\rm cm^{-2}}$ and a photon index of $\Gamma = 2.05\pm0.12$,
with no obvious change in the index within the X-ray PWN.
The measured X-ray absorption indicates that
the distance to the source is much less than $10~{\rm kpc}$ inferred by radio data.
Aside from the PWN, no extended emission was observed around PSR~J2021+3651
even by \textit{Suzaku}.
Archival data from the \textit{XMM-Newton} were also analyzed to complement the \textit{Suzaku} observations,
indicating that the eastern region of the X-ray PWN has 
a similar spectrum ($N(\rm H)=(7.5 \pm 0.9) \times 10^{21}~{\rm cm^{-2}}$ and 
$\Gamma=2.03 \pm 0.10$)
and source extent up to at least $12'$ along the major axis.
The lack of significant change in the photon index 
and the source extent in X-ray are used to constrain the advection velocity or the diffusion coefficient
for accelerated X-ray-producing electrons.
A mean magnetic field of ${\sim}3~\mu{\rm G}$ is required to account for the
measured X-ray spectrum and reported TeV $\gamma$-ray spectrum.
A model calculation of synchrotron radiation and inverse Compton scattering
was able to explain ${\sim}80\%$ of the reported TeV flux,
indicating that the X-ray PWN is a major contributor of VER~J2019+368. 
\end{abstract}




\section{Introduction} \label{sec:intro}


Star-forming regions host several possible cosmic-ray (CR) accelerators such as
supernova remnants (SNRs), pulsars and pulsar wind nebulae (PWNs), Wolf--Rayet stars, and OB associations.
Cygnus-X \citep{Piddington1952,Uyaniker2001}
is one of such nearby star-forming region; it is located at approximately 1.5~kpc
\citep[][]{Rygl2012}
and has long been studied at various wavebands,
although care must be taken
to properly associate individual sources to Cygnus-X,
as there are several spiral arms in the same direction.
A survey of the Northern Hemisphere sky by the Milagro Gamma-Ray Observatory
identified several bright and extended TeV $\gamma$-ray sources \citep{Abdo2007}. 
MGRO~J2019+37 is the brightest Milagro source in the direction of Cygnus-X
with a measured flux of approximately $80\%$ of the Crab Nebula flux at 20~TeV.
Despite extensive studies at various wavebands, the nature of MGRO J2019+37
remained unsettled because of its large source extent 
\citep[$\sigma=0\fdg7$ when modeled with a two-dimensional Gaussian probability density function;][]{Abdo2012}.
The imaging atmospheric Cherenkov telescope array VERITAS carried out a deep observation
of the MGRO~J2019+37 region and resolved it into two sources. The brighter source,
VER~J2019+368, is an extended source that accounts for the bulk of
MGRO~J2019+37 in terms of morphology and spectrum \citep{Aliu2014}.
Its peak is located at right ascension RA (J2000) ${\rm 20^{h}19^{m}25^{s}}$
and declination DEC (J2000) ${\rm 36\arcdeg48'14''}$, and its angular extension is estimated to be
${\sim}0\fdg34$ and ${\sim}0\fdg13$ along its major and minor axis respectively,
with the orientation of the major axis $71\arcdeg$ east of north. 
Its TeV spectrum is hard and represented by a single power-law with a photon index $\Gamma \sim 1.75$
and an integrated energy flux at 1--10~TeV of ${\sim}6.7\times 10^{-12}~{\rm erg~s^{-1}~cm^{-2}}$.
The emission region contains
the energetic pulsar PSR J2021+3651 
\footnote{
The pulse period and its derivatives are $104~{\rm ms}$ and $9.57 \times 10^{-14}$, respectively
\citep{Abdo2009}, giving the surface magnetic field of $3.2 \times 10^{12}~{\rm G}$.
}
with the characteristic age of 17.2~kyr and
spin-down luminosity of $3.4 \times 10^{36}~{\rm erg~s^{-1}}$ \citep{Roberts2002}, its PWN, 
the {H\footnotesize\,II} region Sharpless 104 (Sh~2-104), 
and the Wolf--Rayet star WR~141,
which are potential counterparts of the observed TeV emission.
\textit{Fermi}-LAT \citep{Abdo2009} detected emissions from PSR~J2021+3651 and gave an upper limit
on the extended GeV $\gamma$-ray emissions around it, consistent with the hard spectrum of
VER~J2019+368 in the TeV band. Although the distance to the pulsar is inferred from the radio data
to be ${\ge}10~{\rm kpc}$ \citep{Roberts2002}, this conclusion remains controversial
\citep[e.g.,][]{Etten2008}
and does not coincide with a detailed study of the distance by \citet{Kirichenko2015}.

Among the source classes of possible counterparts mentioned above, 
only PWNs are the established extended TeV $\gamma$-ray sources.
Nevertheless, the association of the X-ray PWN \citep[named G75.2+0.1;][]{Hessels2004}
to VER~J2019+368 is a matter of debate, as its position is offset from the peak in TeV
by ${\sim}20'$ and its reported extent is
$\le15'$ \citep{Roberts2008}, which is much smaller than the size of the TeV $\gamma$-ray emission region.
To resolve this, we carried out deep X-ray observations 
using the X-ray Imaging Spectrometer (XIS) \citep{Koyama2007} on board the \textit{Suzaku}
satellite \citep{Mitsuda2007}, which is very sensitive to extended X-ray emission.
We aimed to accurately measure the spectral and morphological properties of
the X-ray PWN and to observe unknown extended X-ray emissions in the region of
VER~J2019+368. We also analyzed archival \textit{XMM-Newton} \citep{Jansen2001} data 
in order to complement the \textit{Suzaku}-XIS observations,
which did not cover the entire PWN.
This paper is organized as follows.
We describe the X-ray observations and our data reduction in Section~2. 
The results of the data analysis are presented in Section~3, 
in which we provide the
detailed spectral and morphological properties of the X-ray PWN to the west of the pulsar
and make a comparison between the eastern and western regions of the PWN.
Discussion of the PWN's association with VER~J2019+368 based on 
its X-ray properties and a multiwavelength spectrum
is provided in Section~4.
A summary of this study and future prospects are presented in Section~5.

\clearpage

\section{Observations and Data Reduction}

In November 2014, we carried out deep X-ray observations 
of VER~J2019+368 region using \textit{Suzaku}-XIS.
In order to constrain the X-ray properties of the PWN around PSR~J2021+3651
and search for unknown extended X-ray emissions, we conducted two observations.
As shown in Figure~1, these covered the main region of the TeV emission.
The objective of the first pointing (S1) was to characterize the X-ray properties of the 
western region of the PWN,
while the second pointing (S2) had the objective of searching for unknown extended X-ray emissions.

The observations were carried out using the XIS
on the focal plane of the X-Ray Telescope \citep[XRT;][]{Serlemitsons2007} on {\it Suzaku}. 
The XIS consists of two front-illuminated (FI) X-ray charge coupling devices (CCDs) (XIS0 and 3) 
\footnote{Because of an anomaly that occurred in November 2006, the operation of 
another FI sensor, XIS2, has been terminated.}
and one backside-illuminated (BI) X-ray CCD (XIS1). 
The combined XIS and XRT system are sensitive within the energy range of
0.3--12~keV. 
Although its angular resolution is moderate (half-power diameter $\sim 2'$),
the XIS+XRT system provides a low and stable instrumental background 
\citep{Mitsuda2007,Tawa2008} and is, therefore, suitable for the detailed study of extended
emissions
with low surface brightness. 
Data were analyzed using the HEASOFT
\footnote{\url{http://heasarc.nasa.gov/lheasoft/}}
6.15.1 software package with the calibration database
released on October 10, 2015.
We analyzed so-called cleaned events that had passed the following standard event selection criteria: 
(a) Only {\it ASCA}-grade 0, 2, 3, 4, and 6 events were accumulated with hot and
flickering pixels removed.
(b) More than 436~s had elapsed since passing through the South Atlantic 
Anomaly.
(c) The pointing directions were at least $5^{\circ}$ and $20^{\circ}$ above the rim of the Earth 
during the night- and daytime, respectively. 
To further reduce the non-X-ray background (NXB), we also required that 
(d) the geomagnetic cutoff rigidity exceeded 6~GV. 
Details concerning the observation and net exposures of the screened 
events are summarized in Table~1.

In order to complement the \textit{Suzaku} observations which did not cover the entire PWN (see Figure~1),
we also analyzed archival \textit{XMM-Newton} data pointing at the position of PSR~J2021+3651.
\textit{XMM-Newton} is equipped with two types of X-ray CCD, PN and MOS --- both of which have sensitivities within 0.15--12~keV
when combined with the X-ray telescope.
Although the background is rather high and unstable owing to satellite's highly
elliptical orbit, \textit{XMM-Newton} has a larger field of view (FOV) and effective area
and is, therefore, complementary to
\textit{Suzaku}-XIS.
Because PN was operated in timing mode to study the PSR~J2021+3651,
which is not suitable for studying
PWNs, we used only the MOS data.
Among two MOS CCD cameras, we focused on MOS2 data since one CCD chip of MOS1 
that covers part of the PWN
was not functional in this observation.
The SAS
\footnote{\url{https://www.cosmos.esa.int/web/xmm-newton/download-and-install-sas}}
15.0.0 and ESAS
\footnote{\url{http://heasarc.gsfc.nasa.gov/docs/xmm/xmmhp_xmmesas.html}}
13 software packages 
were used in analyzing the data.
Details of the procedure for reducing and estimating the
particle-induced background is given in Section~3.2.
A summary of the \textit{XMM-Newton} observation and net exposure is shown in Table~1.

\clearpage

\begin{figure}[htbp]
\figurenum{1}
\centering
\includegraphics[width=0.5\textwidth]{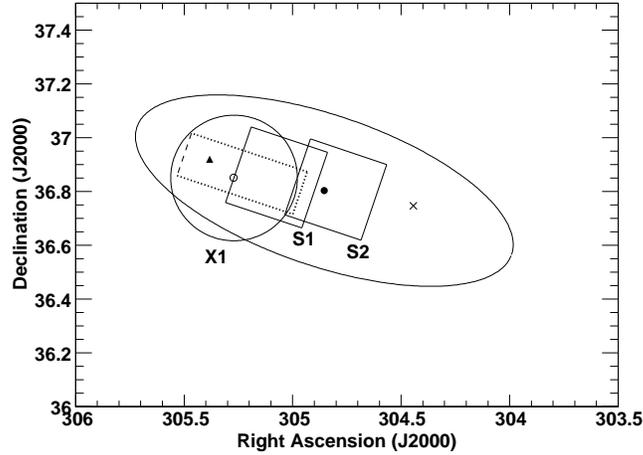}
\caption{
A finding chart of objects and observations. The ellipse shows the TeV emissions of VER~J2019+368
($2.15 \times \sigma$ in both major and minor axes, which includes 90\% of emissions 
if the source distribution is two-dimensional Gaussian). The small filled circle, open circle, triangle, and cross
show the peak position of TeV emission, PSR~J2021+3651, WR~141 and Sh~2-104, respectively.
The two squares show the \textit{Suzaku}-XIS FOV
(the first pointing was located on the eastern side), and the circle shows the FOV of \textit{XMM-Newton} MOS.
The dotted rectangle represents the measured size of the X-ray PWN,
with the eastern side is shown by a dashed line since the source extent was not constrained (see Section~3).
}
\end{figure}

\begin{table}[htbp]
\caption{Summary of observations}
\begin{center}
\small
\begin{tabular}{cccccccc}
\hline\hline
Observatory     & Region & \multicolumn{2}{c}{Pointing\tablenotemark{a}} & Observation date & Net exposure \\
                &        & RA (deg) & DEC (deg)                          &                  & (ks) \\
\hline
\textit{Suzaku} & S1 & 305.064 & 36.873 & 2014 Nov. 09 & 35.0 \\
                & S2 & 304.792 & 36.828 & 2014 Nov. 10 & 35.7 \\
\textit{XMM-Newton} & X1 & 305.273 & 36.851 & 2012 Apr. 07 & 83.4 \\
\hline
\end{tabular}
\tablenotetext{a}{
Position of the center of the XIS (\textit{Suzaku}) or MOS (\textit{XMM-Newton}) FOV.
}
\end{center}
\end{table}

\clearpage

\section{Data Analysis and Results}
\subsection{Suzaku Data}

\subsubsection{X-ray Images and PWN-West Morphology}

We extracted X-ray images from XIS3 (FI CCD), which has 
better imaging quality than BI CCD (XIS1) owing to its
lower instrumental background \citep{Mitsuda2007,Tawa2008}. 
Although XIS0 also has good imaging quality, we did not use it to construct an image
in order to avoid artifacts resulting from its unusable area ({$\sim$}1/4 of the CCD chip).
We defined the soft and hard bands as 0.7--2 and 2--10~keV, respectively, 
and excluded the corners of the CCD chips illuminated by the ${\rm ^{55}Fe}$ calibration sources. 
We then estimated the NXB contribution from the nighttime Earth data and subtracted it from the images using 
{\tt xisnxbgen} \citep{Tawa2008}.
Vignetting was then corrected by dividing the soft- and hard-band images by flat sky images simulated at 
1 and 4~keV, respectively, using the XRT+XIS simulator {\tt xissim} \citep{Ishisaki2007}. 
In the flat image simulations, we assumed a uniform intensity of ${\rm 1~photon~s^{-1}~cm^{-2}~sr^{-1}}$,
and therefore, the approximate unit of the obtained vignetting-corrected image is 
${\rm photons~s^{-1}~cm^{-2}~sr^{-1}}$.
The obtained images are shown in Figure~2, in which 
smoothing with a Gaussian kernel of $\sigma=0.\hspace{-2pt}'28$ is applied for visualization.

Extended emission from the western part of the PWN is apparent in our first observation (S1, Figure~1),
but no obvious extended emission is seen in our second observation (S2, Figure~1).
We also identified two bright sources in S1: PSR~J2021+3651, located at the east edge of the CCD chip,
and a bright field star, USNO-B1.0 1268-044892 
\citep[already reported by][]{Etten2008}, which is seen mainly in the soft band.
The PWN emission is roughly along the south side of the CCD that was tilted by $71\fdg4$
east from the north, suggesting that the major axis of the X-ray PWN is almost parallel to
that of VER~J2019+368.
To examine the source extent quantitatively, we defined $5' \times 1'$ rectangles
\footnote{In 15th to 17th regions (from the pulsar) in S1 observation,
we used $4' \times 1'$ (instead of $5' \times 1'$) rectangles 
to avoid the corners illuminated by the ${\rm ^{55}Fe}$ calibration sources.}
as shown in Figure~2 and calculated the PWN count rate profile.
We used both XIS0 and XIS3 to conduct a morphology analysis of the PWN emission along its major axis.
The size and position of the rectangles were chosen to avoid the unusable area of XIS0.
We also removed two sources (PSR~J2021+3651 and USNO-B1.0 1268-044892) in S1 and one hard source
(presumably the background active galactic nucleus) in S2:
the radius of the circles for exclusion was $120''$, $90''$, and $90''$ for
PSR~J2021+3651, USNO-B1.0 1268-044892, and the hard source seen in S2, respectively.
As PSR~J2021+3651 was located near the edge of the CCD chip,
its position was not discernible from the XIS image.
As the position accuracy of the XIS is known to be 
${\sim}20''$ \citep{Uchiyama2008}, we did not use the reported position of the pulsar;
instead, we referred to USNO-B1.0 1268-044892 and obtained the shifts as $-22''$ and $-9''$ 
in RA and DEC, respectively,
and determined the position of PSR~J2021+3651 in our image.
From the position of PSR~J2021+3651 toward the southwest 
(with the orientation of $108\fdg6$ west from the north), 
we defined 17 rectangles in S1 and 11 rectangles in S2,
with three rectangles overlapped. We then examined the morphology of the PWN up to $25'$ 
from the position of the pulsar
after first subtracting the NXB estimated by {\tt xisnxbgen} \citep{Tawa2008}.
The remaining X-ray background --- presumably the cosmic X-ray background (CXB) and the 
Galactic ridge X-ray emission (GRXE) \citep[e.g.,][]{Worrall1982,Warwick1985,Koyama1986},
which are
expected to be almost uniform within the XIS FOV ---
was estimated using $10' \times 4'$ rectangles in each observation as shown in Figure~2. 
The NXB-subtracted background count rate was
subtracted from the NXB-subtracted source count rate
with vignetting taken into account.
Finally, the obtained count rate of each bin was corrected for vignetting and the region size
(normalized to the count rate of the ninth rectangle from the pulsar
which is the closest to the FOV center of the S1 observation),
as summarized in Figure~3.
It is seen from the figure that the PWN emission extends from the pulsar
in the southwest direction by up to $15'$ in the soft band and $18'$ in the hard band. 
Although we corrected for the vignetting effect,
it is severe at high energies and near the edge of the XIS \citep{Serlemitsons2007}.
Therefore, we concluded rather conservatively that the PWN emission extends from the pulsar
in the southwest direction at least up to $15'$ in both the soft and hard bands.

We also examined the PWN morphology in the minor axis direction,
and in the region between $3'$ and $6'$ from the pulsar along the major axis 
to avoid contamination
from the pulsar, as shown in Figure~4(a).
Because the XIS0 has an unusable area on the south side of the FOV,
we only used XIS3 and the hard band in order to avoid emissions from
USNO-B1.0 1268-044892.
Background subtraction and vignetting correction
were conducted in the same manner as in the morphology study along the major axis.
The obtained count rate profile is shown in Figure~4(b),
in which distance is measured from the south edge of the XIS toward the north, 
and the bins from $3'$ to $8'$ are located
within rectangles used to study the morphology along the major axis.
It is seen from the figure that
the PWN emission has a source extent of at least $10'$ 
along the minor axis;
thus, these results show for the first time that 
the western region of the PWN has a 
source extent of
at least $15'$ and $10'$ along the
major and minor axis, respectively.
The count rates in the $3'$--$8'$ bins (i.e., those within the area
of study along the major axis)
and within $0'$--$10'$ bins (the entire PWN emission) are $12.64\pm0.74$ 
and $18.60\pm1.02~{\rm c~s^{-1}}$, respectively,
giving a ratio of $1.47 \pm 0.12$ to convert the flux within a region of 
$5'$ width to that of the entire PWN-West emission (see also Section~3.3).

\clearpage

\begin{figure}[ht!]
\figurenum{2}
\gridline{
\fig{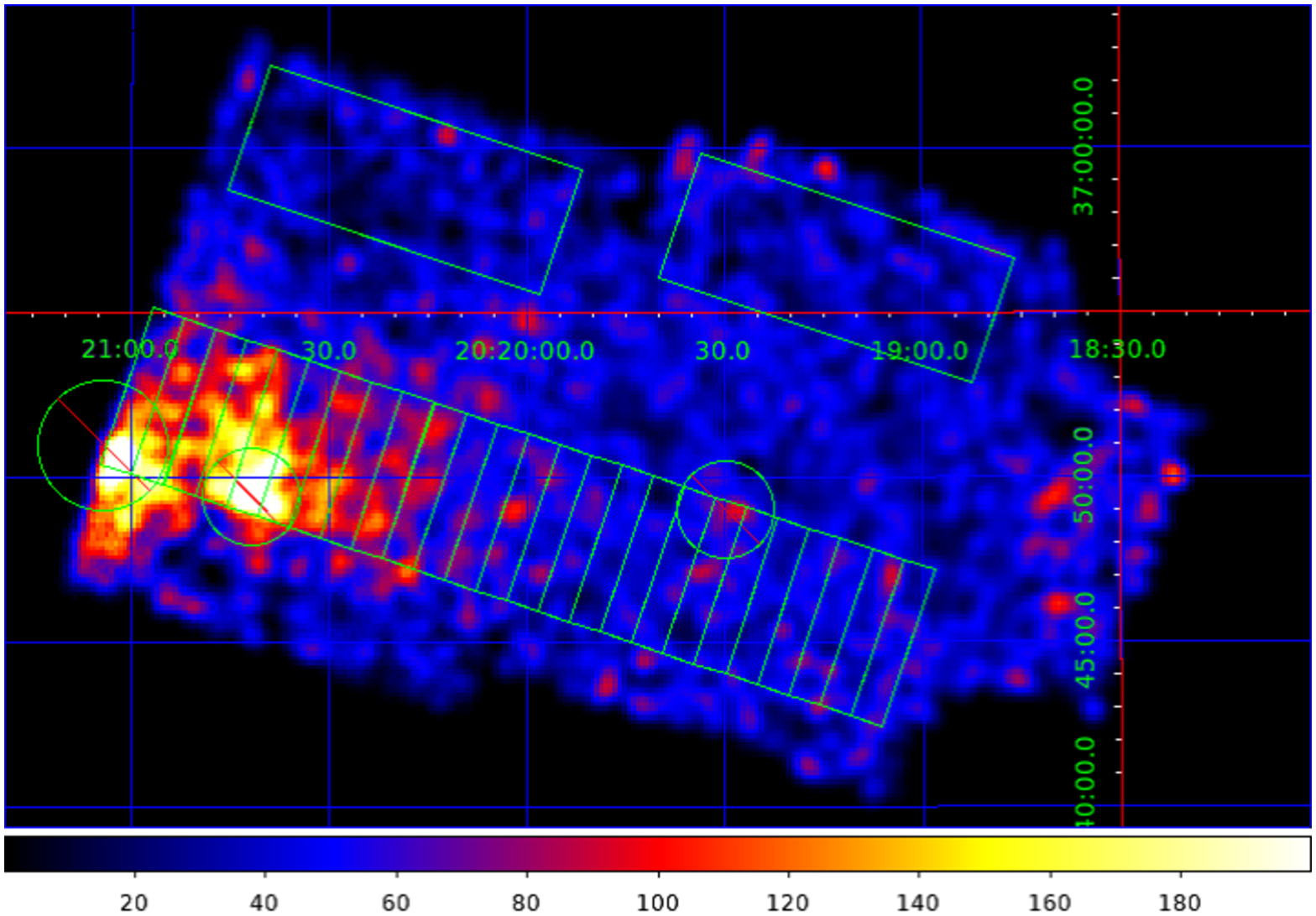}
{0.5\textwidth}{(a)}
\fig{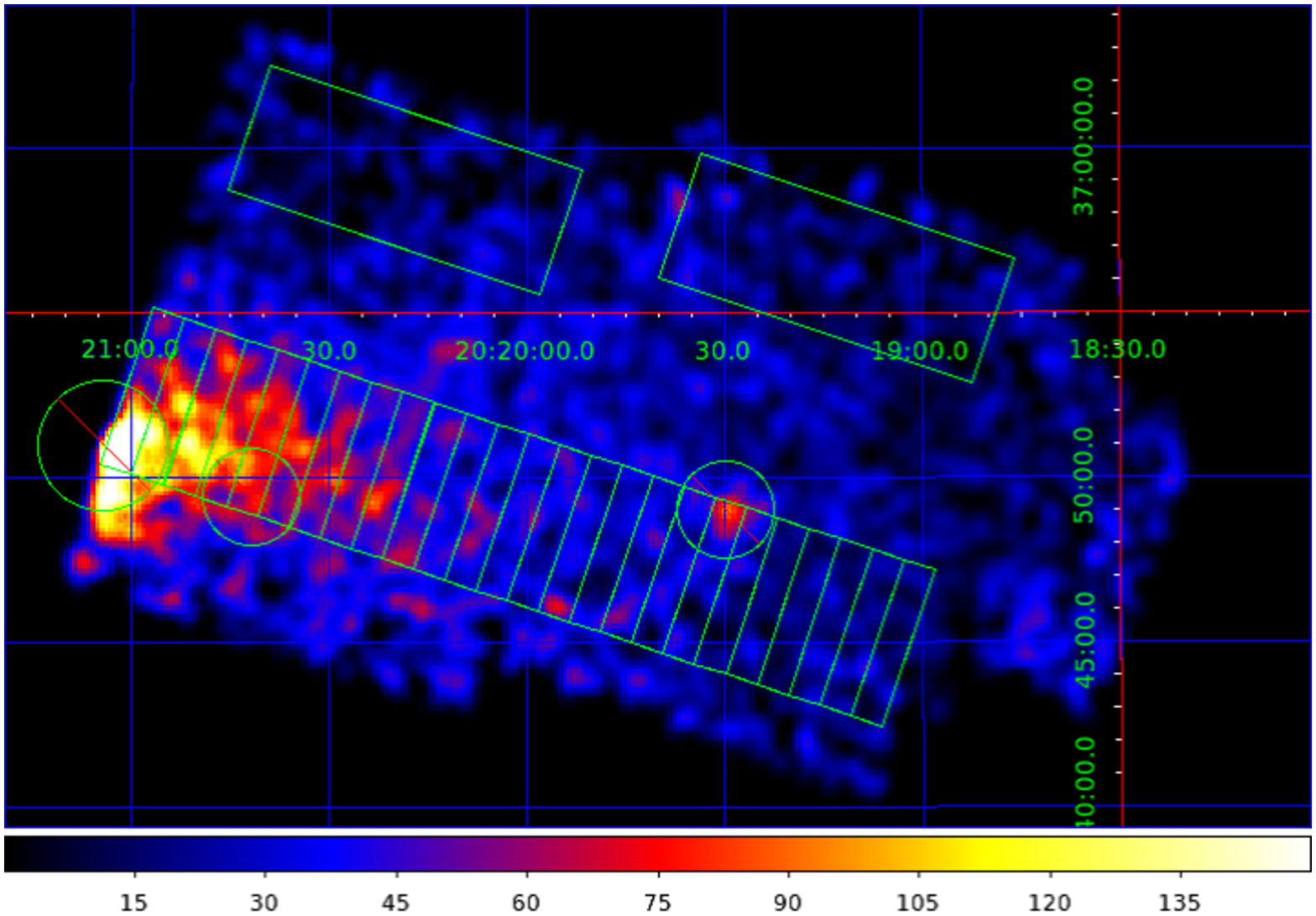}
{0.5\textwidth}{(b)}
}
\caption{
X-ray intensity maps in equatorial coordinates, in which S1 and S2 observations are combined
(left and right halves of the image correspond to S1 and S2 observations, respectively).
(a) Soft-band (0.7--2~keV) image.
(b) Hard-band (2--10~keV) image.
The unit is approximately ${\rm photons~s^{-1}~cm^{-2}~sr^{-1}}$ in each energy band (see text).
Rectangles used to study the morphology along the major axis of the PWN and estimate the background are overlaid
The circles exclude contamination from point sources.
\label{fig:f1}
}
\end{figure}

\begin{figure}[ht!]
\figurenum{3}
\gridline{
\fig{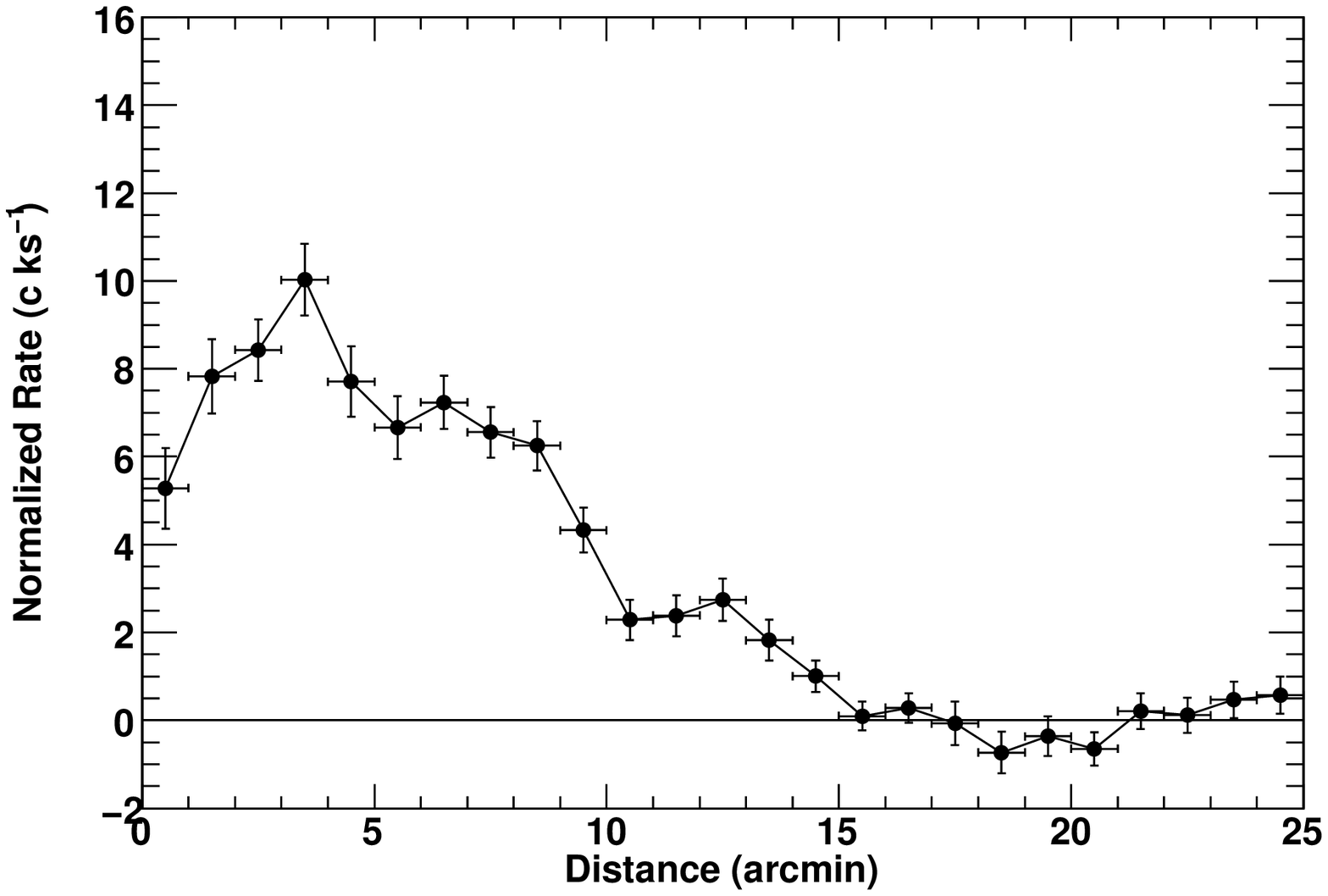}
{0.5\textwidth}{(a)}
\fig{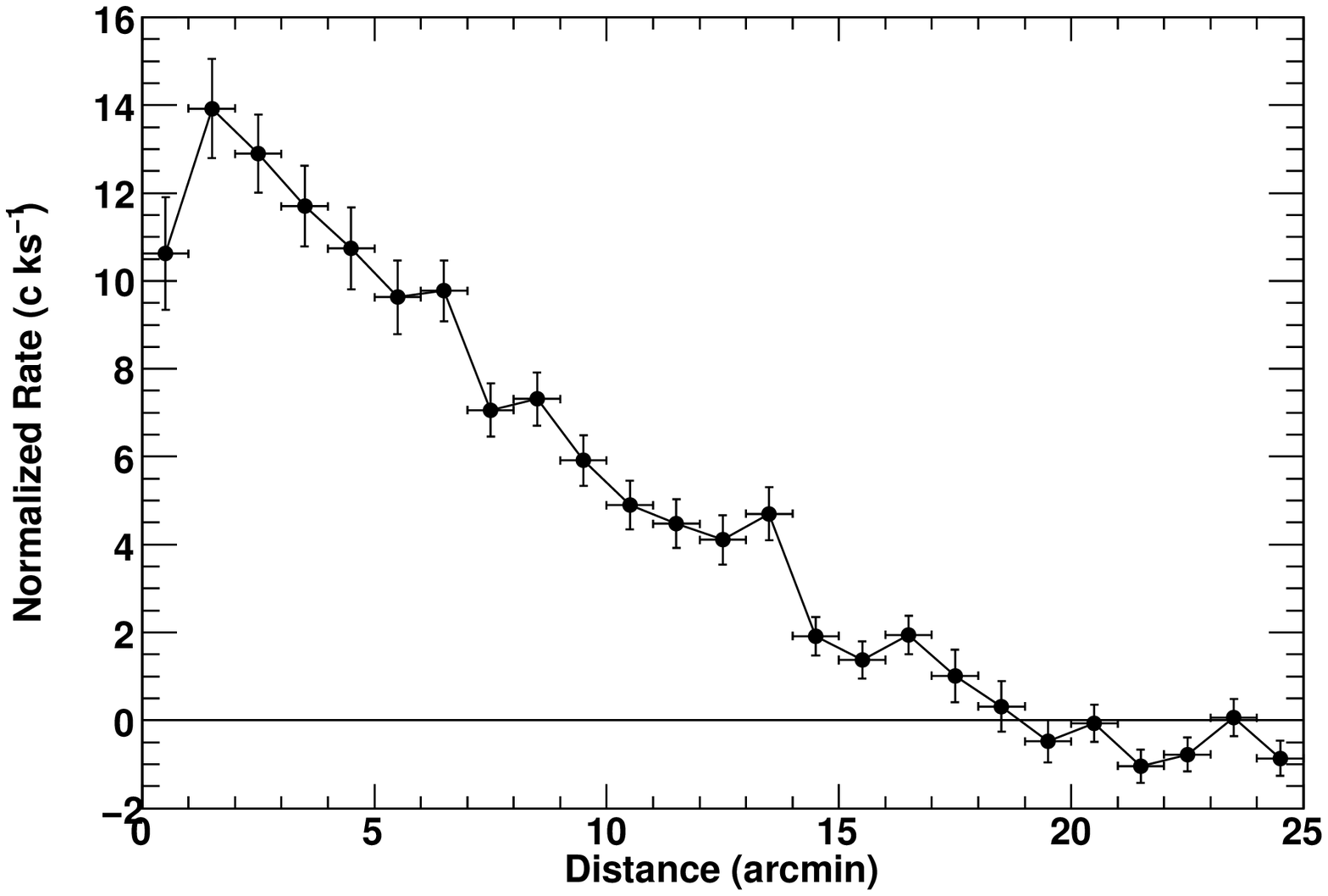}
{0.5\textwidth}{(b)}
}
\caption{
(a) Soft-band count rate along the PWN major axis.
(b) Hard-band count rate along the PWN major axis.
The distance is measured from the eastern edge of the XIS of the S1 observation toward the west and is nearly the same as
the distance from the pulsar.
In both panels, NXB was subtracted and the X-ray background was
subtracted with vignetting taken into account.
The count rate of each bin was normalized (with vignetting taken into account) to that in the ninth bin
which is the closest to the FOV center of the S1 observation.
\label{fig:f1}
}
\end{figure}

\begin{figure}[ht!]
\figurenum{4}
\gridline{
\fig{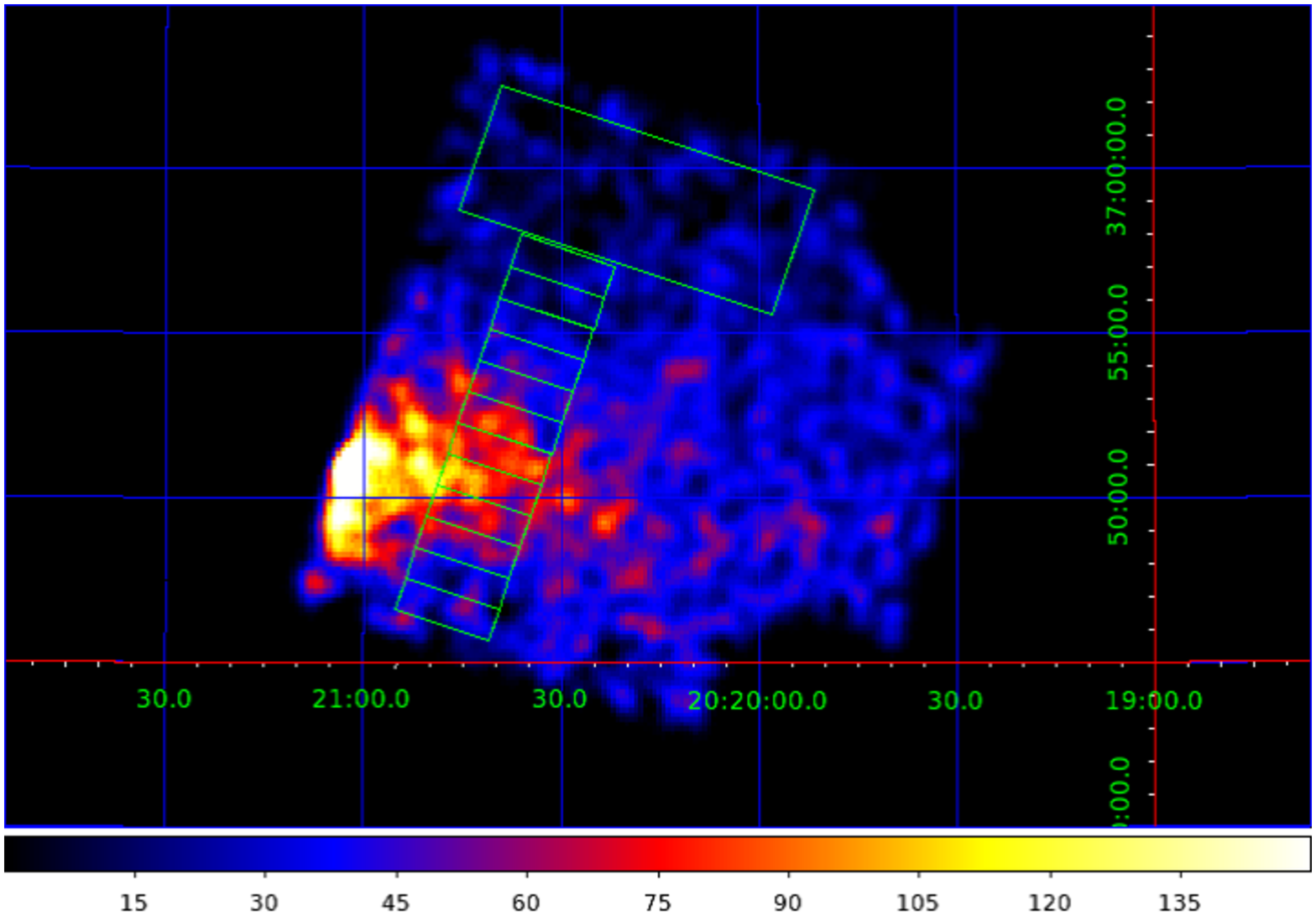}
{0.5\textwidth}{(a)}
\fig{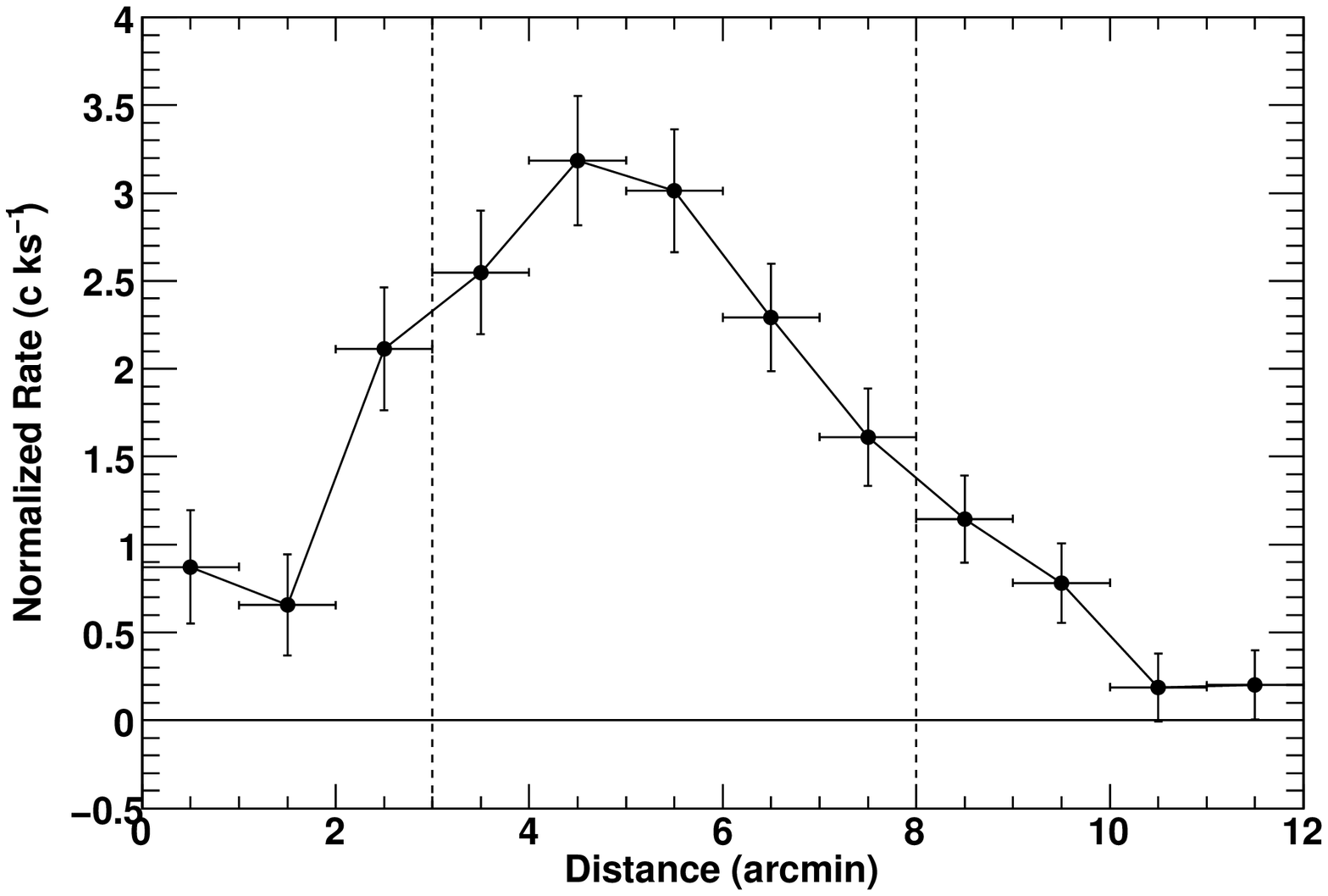}
{0.5\textwidth}{(b)}
}
\caption{
(a) A 2--10~keV intensity map of S1 observation, 
with rectangles for studying the morphology overlaid along the minor axis.
(b) Background-subtracted and vignetting-corrected count rate along the minor axis.
The count rate of each bin is normalized (with vignetting taken into account) to that in $5'$--$6'$.
Vertical dotted lines indicate the area used to study the count rate profile along the major axis.
\label{fig:f1}
}
\end{figure}

\clearpage

\subsubsection{Spectrum of the PWN-West}

As described in Section ~3.1.1,
we confirmed that the PWN-west region extends up to (at least) $15'$ westward from the pulsar.
We then extracted spectra obtained by three CCD cameras (XIS0, XIS1, and XIS3)
for 15 rectangles from S1, starting with the rectangle closest to the pulsar
and with two point sources excluded,
in order to maximize the photon statistics while avoiding the unusable area of XIS0. 
On the basis of the morphology of the major axis shown in Figure~3,
we assumed a linear decrease in intensity (from 1 to 0 in relative) from $0'$ to $15'$ in calculating 
the ancillary response files (ARFs)
using {\tt xissimarfgen} \citep{Ishisaki2007}.
The losses of effective area owing to the exclusion of point sources
and area illuminated by calibration sources were taken into account in calculating the ARFs.
In the spectral analysis, the response matrix files (RMFs) were calculated using {\tt xisrmfgen}, 
and the integrated NXB spectrum over the source spectrum region
was estimated using {\tt xisnxbgen} \citep{Tawa2008}
and subtracted from the source spectrum.
As the NXB-subtracted X-ray spectrum was expected to suffer from the (X-ray) background
owing to the CXB and GRXE,
the background was estimated again using the $10' \times 4'$ source-free region (see Figure~2).
We first subtracted the NXB contribution from the background spectrum
and then subtracted the NXB-subtracted background spectrum from the NXB-subtracted source spectrum
with vignetting at 2~keV taken into account.
The vignetting correction factors at 1 and 4~keV differ from that at 2~keV by only
${\sim}3\%$ and ${\sim}2\%$, respectively.
The obtained spectrum was well fitted
[reduced chi-square $\chi^{2}/{\rm degrees~of~freedom~(DOF)} = 211.1/188$]
by an absorbed power-law model
(${\tt wabs} \times {\tt pow}$ in {\tt XSPEC}),
as shown in Figure~5 and Table~2.
The best-fit hydrogen column density of the photoelectric absorption, $N({\rm H}) = (8.2^{+1.3}_{-1.1}) \times 10^{21}~{\rm cm^{-2}}$, 
\footnote{Here and hereafter, errors are calculated for single-parameter 90\% confidence limit.}
was consistent with the absorption toward the vicinity of the pulsar
measured by the \textit{Chandra}
X-ray Observatory as reported by \citet{Etten2008}
($6.7^{+0.8}_{-0.7} \times 10^{21}~{\rm cm^{-2}}$). 
Therefore, we confirmed that the extended X-ray emission
comes from the PWN around PSR~J2021+3651.
In order to examine the possible spectral change along the major axis,
we divided the source region into five segments,
each $3'$ length. We repeated the same analysis procedure
[response calculation, NXB subtraction, 
and X-ray background subtraction with vignetting correction]
as described above and fit each of the spectra
with an absorbed power-law model.
We first let the absorption free to vary in each region and obtained the parameters as summarized in Table~2.
Although there seems to be a slight softening of the spectra in outer regions (${\rm distance \ge 9'}$), 
we observe a correlated increase of the absorption 
and the photon indices of all five subregions are consistent with that of the whole spectrum within statistical errors.
We also fixed the absorption at $8.2 \times 10^{21}~{\rm cm^{-2}}$,
which was the best-fit value of the whole spectrum.
As is seen from Table~2, the photon index does not change significantly over the 
entire western region of the PWN.
We can also see that the intensity gradually decreases in a manner 
approximately proportional to the distance from the pulsar.

\clearpage

\begin{figure}[ht!]
\figurenum{5}
\centering
\includegraphics[height=0.6\textwidth,angle=-90]{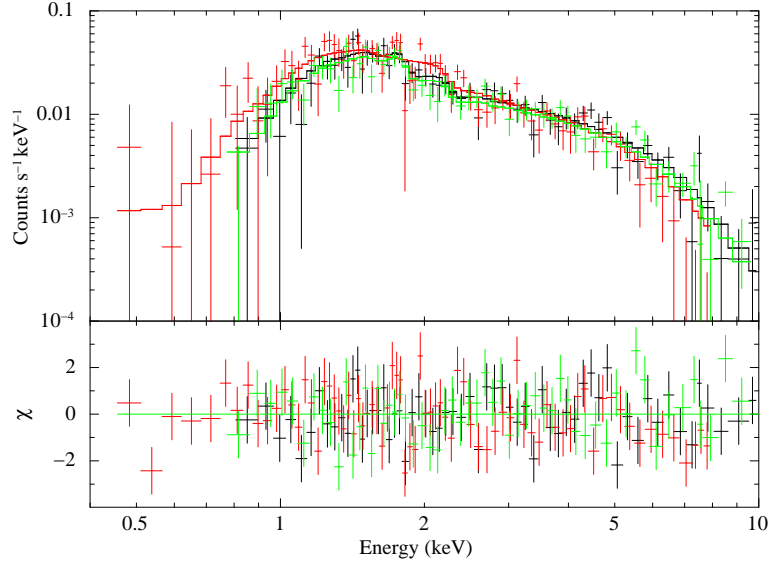}
\caption{
PWN-west spectrum (crosses) obtained by \textit{Suzaku}-XIS integrated over
$15'$ (along the major axis) $\times$ $5'$ (along the minor axis except for the furthermost part),
fitted by an absorbed power-law model (lines). Black, red, and green crosses indicate the XIS0,
XIS1, and XIS3 data, respectively. The bottom panel shows the residuals.
\label{fig:f1}
}
\end{figure}

\begin{table}[htbp]
\caption{Summary of the spectral fits of the PWN-west by \textit{Suzaku}-XIS}
\begin{center}
\small
\begin{tabular}{cccccc}
\hline\hline
Region$^{a}$ & $N({\rm H})$                    & $\Gamma$ & $f({\rm 0.5\mbox{--}2~keV})$ & $f({\rm 2\mbox{--}10~keV})$ & $\chi^{2}$/DOF \\
       & ($10^{21}~{\rm cm^{-2}}$) &          & ($10^{-13}~{\rm erg~s^{-1}~cm^{-2}}$) & ($10^{-13}~{\rm erg~s^{-1}~cm^{-2}}$) & \\ 
\hline
all($0'\mbox{--}15'$)    & $8.2^{+1.3}_{-1.1}$ & $2.05\pm0.12$ & $6.04^{+0.42}_{-0.40}$ & $26.1^{+1.6}_{-1.8}$ & 211.1/188 \\ 
\hline
$0'\mbox{--}3'$ & $8.2^{+2.1}_{-2.0}$ & $2.07\pm0.21$ & $1.97^{+0.16}_{-0.22}$ & $8.27^{+0.89}_{-1.06}$ & 65.5/61 \\
$3'\mbox{--}6'$ & $6.2^{+1.7}_{-1.5}$ & $1.96\pm0.18$ & $1.68^{+0.17}_{-0.20}$ & $6.52^{+0.76}_{-0.73}$ & 60.5/67 \\
$6'\mbox{--}9'$ & $7.1^{+1.8}_{-1.5}$ & $2.06\pm0.18$ & $1.22\pm0.13$ & $4.71^{+0.51}_{-0.52}$ & 128.9/99 \\
$9'\mbox{--}12'$ & $12.5^{+3.9}_{-3.2}$ & $2.30\pm0.32$ & $0.57^{+0.07}_{-0.12}$ & $2.77^{+0.39}_{-0.53}$ & 85.3/73 \\
$12'\mbox{--}15'$ & $13.7^{+5.8}_{-4.4}$ & $2.29\pm0.42$ & $0.30^{+0.07}_{-0.10}$ & $1.69^{+0.35}_{-0.61}$ & 54.7/52 \\ 
\hline
$0'\mbox{--}3'$ & 8.2(fixed) & $2.07\pm0.10$ & $1.96^{+0.20}_{-0.18}$ & $8.26^{+0.86}_{-0.90}$ & 65.5/62 \\
$3'\mbox{--}6'$ & 8.2(fixed) & $2.14\pm0.11$ & $1.61^{+0.15}_{-0.18}$ & $6.28^{+0.63}_{-0.61}$ & 64.5/68 \\
$6'\mbox{--}9'$ & 8.2(fixed) & $2.15\pm0.10$ & $1.20^{+0.11}_{-0.12}$ & $4.62^{+0.51}_{-0.46}$ & 130.0/100 \\
$9'\mbox{--}12'$ & 8.2(fixed) & $1.94\pm0.14$ & $0.60^{+0.08}_{-0.09}$ & $2.93^{+0.47}_{-0.46}$ & 90.5/74 \\
$12'\mbox{--}15'$ & 8.2(fixed) & $1.84\pm0.19$ & $0.33^{+0.07}_{-0.08}$ & $1.80^{+0.71}_{-0.36}$ & 59.3/53 \\ 
\hline
\end{tabular}
\tablenotetext{}{Notes: The $N({\rm H})$ and $\Gamma$ are the 
hydrogen column density of the photoelectric absorption and the photon index of the power-law model, respectively.
$f({\rm 0.5\mbox{--}2~keV})$ and $f({\rm 2\mbox{--}10~keV})$ are absorption-uncorrected fluxes in
0.5--2 and 2--10~keV, respectively. Errors are calculated for single-parameter 90\% confidence limit.}
\tablenotetext{a}{The length of the integration region along the major axis is given.
The width of the region along the minor axis is $4'$ (in the length of $14'\mbox{--}15'$) or
$5'$ (elsewhere). See also Figure~2.}
\end{center}
\end{table}

\clearpage

\subsection{\textit{XMM-Newton} Data}

\subsubsection{X-ray Images and PWN Morphology}

In order to examine the overall properties of the PWN, we also analyzed archival data
produced when the \textit{XMM-Newton} was aimed 
at the position of PSR~J2021+3651 (see Table~1).
To reduce the particle-induced background and estimate the residual background
as accurately as possible, we processed data using the ESAS software package.
Details of the \textit{XMM-Newton} CCD-camera background and analysis procedures for extended objects can be found in
\citet{Kuntz2008} and \citet{Snowden2014}.
We first excluded a period of time in which the data was severely contaminated by highly fluctuating background induced
by soft protons, by making a light curve in 2.5--8.5~keV from the entire FOV,
and created a count map.
A cut was made by setting the threshold at $\pm 1.5\sigma$ from the average,
resulting in the net exposure of 83.4~ks. 
We next created a count map due to the quiescent particle background (QPB)
based on data obtained when the filter wheel was in the closed position (FWC data),
subtracted it from the cleaned count map, and then divided the subtracted map by an exposure map to correct the vignetting.
The procedures described above were made by running {\tt mos-filter} and {\tt adapt} commands.
The obtained background-excluded/subtracted and exposure-corrected images in the soft-band (0.7--2~keV) and 
hard-band (2--10~keV) are shown in Figure~6, showing
that the western and eastern regions of
the PWN have a similar source extent.
Three bright sources are also identifiable:
PWN~J2021+3651 in the middle of the image, 
WR~141 to the northeast of the pulsar, 
and USNO-B1.0 1268-044892 to the southwest of the pulsar.

To examine the morphology of the western and eastern regions of the PWN,
we defined 12 rectangles of $5' \times 1'$ in each region
as shown in the figure, similar to those used for the \textit{Suzaku}-XIS data analysis.
We excluded PSR~J2021+3651 using a circular masking region
with a radius of $60''$, and WR~141 and USNO-B1.0 1268-044892 
using circular regions with a radius of $45''$.
We also excluded several less-bright sources as using circular regions of $30''$ radius indicated by the figure.
Although we applied the temporal filtering based on the light curve
to reduce the soft proton contamination and subtracted the QPB,
there remains non-negligible residual background induced by soft protons.
To estimate the residual background, we extracted the spectrum from the entire FOV and fit it with
a model to represent the X-ray emission plus residual soft-proton background modeled
as a simple power-law convolved with the response matrix
of diagonal unity elements \citep{Kuntz2008,Snowden2014}. 
The details of the procedure and best-fit model parameters are given in Appendix~A.
In the following analysis of the morphology and spectrum of the PWN, the spectral index of the 
soft proton contamination is fixed to the best fit value for the entire FOV, and the normalization is scaled by using the
{\tt proton-scale} command.

The X-ray background (presumably the CXB and GRXE)
for the PWN was estimated by calculating the count rate of the background region
($5' \times 4'$ rectangle located in northwest of the pulsar shown in Figure~6) 
after the QPB and the residual soft-proton background estimated from the entire FOV
was subtracted (see above).
We thus obtained the count rate profile of the western and eastern part of the PWN, with 
the QPB contribution estimated based on FWC data and subtracted, 
the residual soft-proton contamination estimated from the entire FOV and subtracted,
and the X-ray background estimated from the background region and subtracted with the vignetting taken into account,
as summarized in Figure~7.
As was done for the morphology analysis by \textit{Suzaku} data (Section~3.1.1), 
the count rate of each bin was corrected for vignetting and the region size 
(normalized to the entire rectangle of $5' \times 1'$ closest to the pulsar in the PWN-west).
It is seen from the figure that the PWN emission extends from the pulsar in the 
southwest and northeast directions
by up to at least $12'$ in both the soft (0.7--2~keV) and hard (2--10~keV) bands. 
We can also see that the count rate profile of the PWN-west is roughly the same
as that seen by the \textit{Suzaku} morphology analysis, while the decrease of the count rate is
less pronounced in the PWN-east.
We also note that the estimated
ratio of the X-ray count rate (PWN, CXB, and GRXE) to the NXB count rate (QPB and residual soft proton background)
at the western edge of the PWN in \textit{XMM-Newton} image
($11'\mbox{--}12'$ away from the pulsar) is about 0.29 in 2--10~keV, while that seen in \textit{Suzaku} data is about 4.3.
Therefore the \textit{XMM-Newton} data might suffer from the larger systematic uncertainty of the NXB. 
This could be 
why we observe an enhancement of the intensity in western and eastern edges of the \textit{XMM-Newton} image
in the hard band (Figure~6b),
the former of which was absent in \textit{Suzaku} image (Figure~2b).

\subsubsection{Spectrum of the PWN}

We then proceed to the spectral analysis. We first extracted spectra for the western and eastern parts
of the PWN for 12 rectangles starting with those closest to the pulsar with bright spots excluded
(see Figure~6). On the basis of the morphology along the major axis shown in Figure~3 and 7,
we assumed a linear decrease in intensity (from 1 to 0 in relative) from $0'$ to $15'$ in calculating the ARFs
using the {\tt arfgen} command for the PWN-west, and another linear decrease in intensity (from 1 to 0.5 in relative)
from $0'$ to $15'$ for the PWN-east. The losses of the effective area owing to the exclusion of 
point sources were taken into account in calculating the ARFs. 
In the spectral analysis, the QPB contribution was estimated using {\tt mos-filter} and subtracted from the source spectrum,
the residual soft-proton contamination was estimated using the data of the entire FOV and subtracted
with the scale factor calculated by {\tt proton-scale}, and 
the contribution from the X-ray background were estimated by 
simultaneously fitting the spectra of
the source and background regions shown in Figure~6. 
The extended PWN emission was modeled by an absorbed power-law model 
(${\tt wabs} \times {\tt pow}$ in {\tt XSPEC}) in the source spectrum.
We also analyzed the spectrum of the so-called "Arc" \citep{Etten2008}
by the same procedure
with a flat intensity profile assumed in calculating the ARF.
The obtained spectra are shown in Figure~8 and parameters of the source spectra
are summarized in Table~3.
Detailed descriptions of the spectral modeling and obtained parameters of the background region are given in Appendix~A.
It is seen from Table~3 that the absorption ($N({\rm H})$), photon index ($\Gamma$), and flux are similar
between the PWN-west and PWN-east, and similar $N({\rm H})$ and $\Gamma$
are obtained for the Arc,
supporting the same physical origin of three regions.
$N({\rm H})$ and $\Gamma$ of the PWN-west are similar to those measured by \textit{Suzaku} (Table~2),
and the obtained flux in 2--10~keV agrees with that integrated over $0'\mbox{--}12'$ by \textit{Suzaku} in ${\sim}10\%$.
We also divided the western/eastern regions into four segments (each $3'$ length) and fitted each of the spectra
as we did for \textit{Suzaku} data and summarize the results in Table~4, 
in which the significant change of the spectral index was not seen.

\clearpage

\begin{figure}[ht!]
\figurenum{6}
\centering
\gridline{
\fig{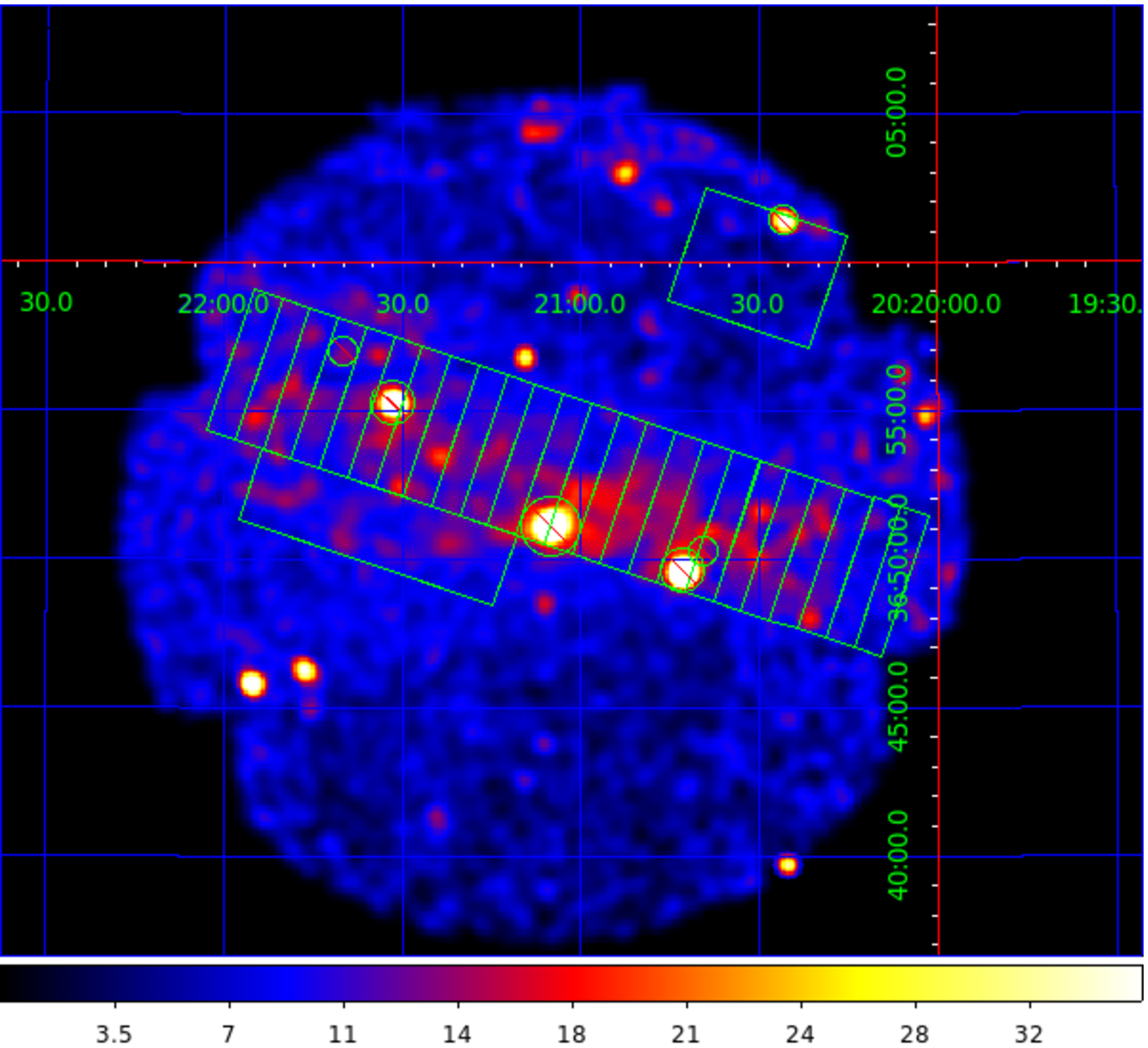}
{0.5\textwidth}{(a)}
\fig{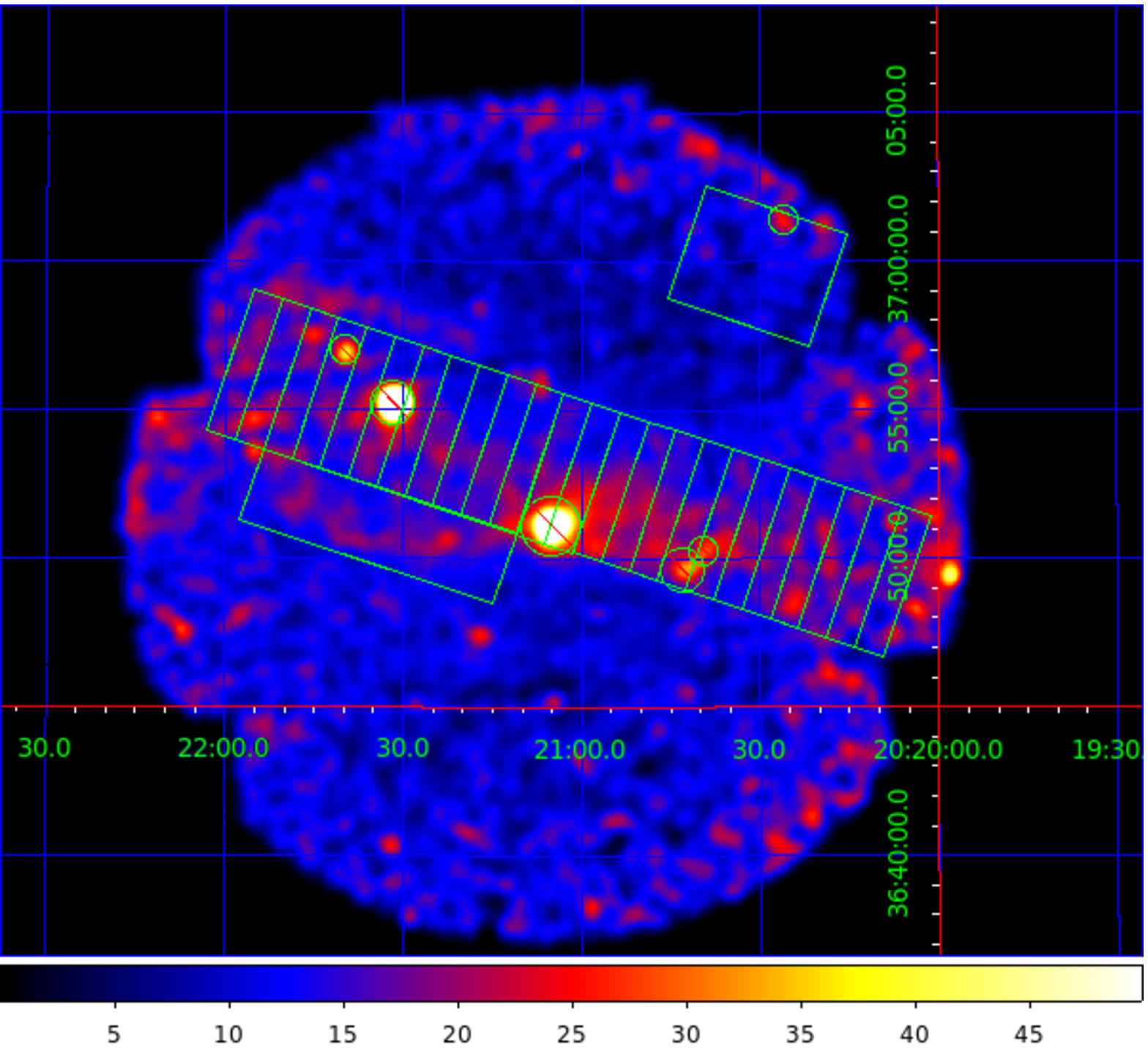}
{0.5\textwidth}{(b)}
}
\caption{
The QPB-subtracted and exposure-corrected
count maps (in the unit of ${\rm counts~s^{-1}~deg^{-2}}$)
around PSR~J2021+3651 taken by \textit{XMM-Newton} MOS with regions for the analysis overlaid.
(a) Soft-band (0.7--2~keV) image. (b) Hard-band (2--10~keV) image.
24 rectangles are for the morphology and spectral analysis of the PWN-west and PWN-east, 
a rectangle of $9' \times 2.\hspace{-2pt}'5$ is for studying
the spectrum of the Arc, and a rectangle of $5' \times 4'$ is a background region.
Smoothing with a Gaussian kernel of $\sigma=0.\hspace{-2pt}'25$ is applied for visualization.
PSR~J2021+3651 (middle in the image), WR~141 (northeast to the pulsar) and USNO-B1.0 1268-044892 (northwest to the pulsar)
are masked using circular regions in the morphology and spectral analysis. 
Several bright spots are also masked with smaller circular regions in the source and background regions.
The circle adjacent to USNO-B1.0 1268-044892 is to exclude bright spot seen in the hard band.
\label{fig:f1}
}
\end{figure}

\begin{figure}[ht!]
\figurenum{7}
\gridline{
\fig{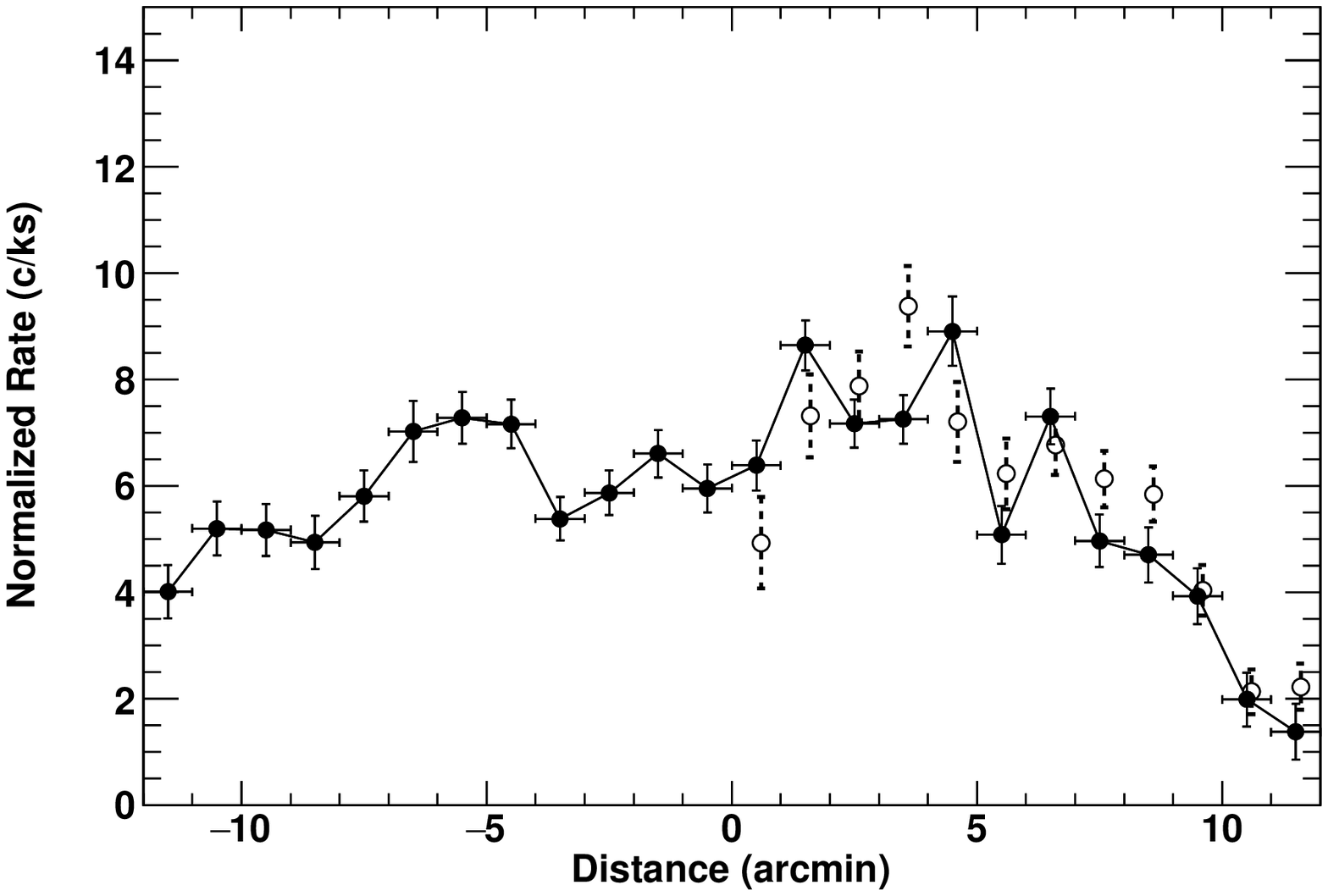}
{0.5\textwidth}{(a)}
\fig{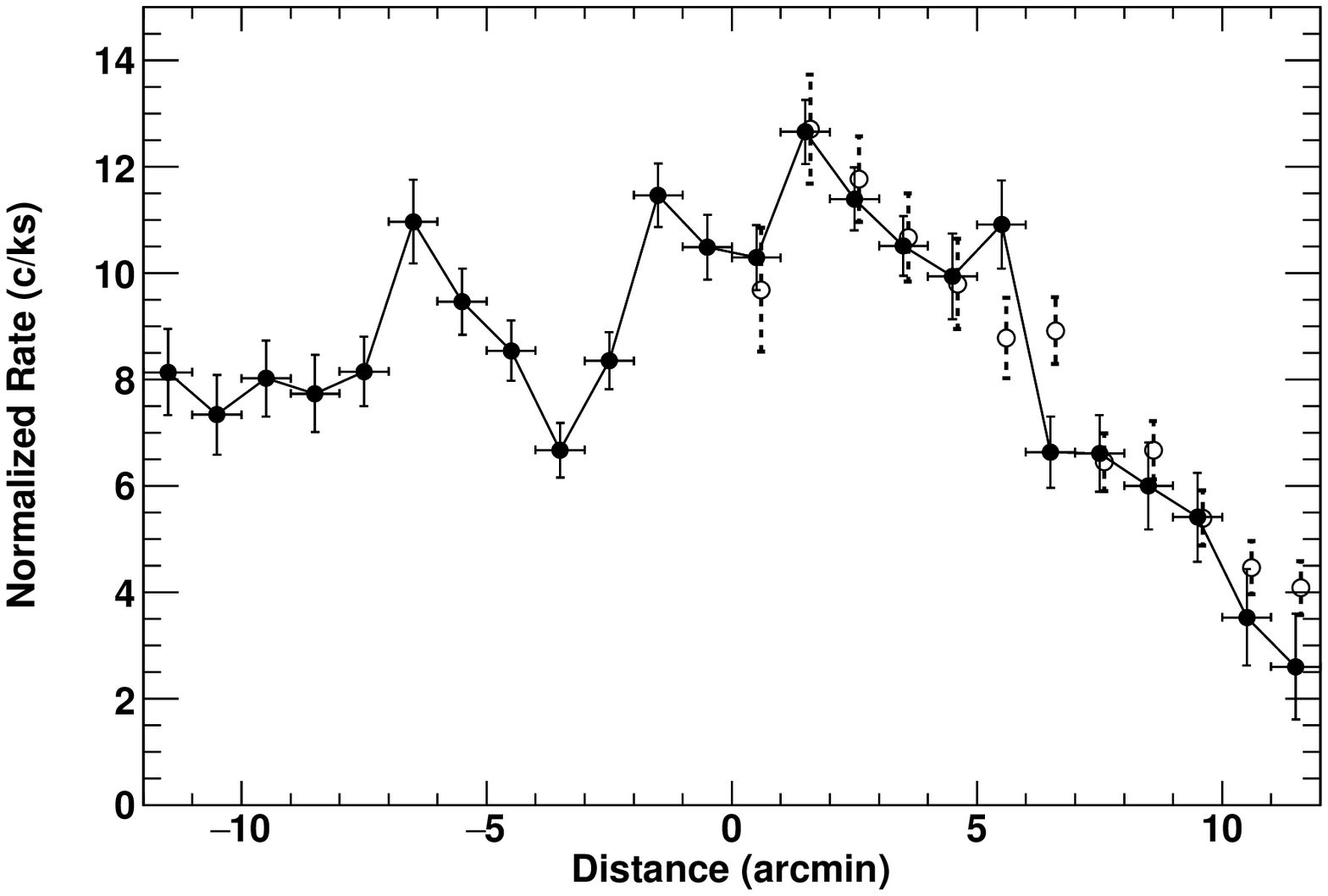}
{0.5\textwidth}{(b)}
}
\caption{
Count rate profiles along the PWN major axis in (a) 0.7--2~keV 
and (b) 2--10~keV.
In the soft-band profile, data in 1.4--1.6 and 1.7--1.8~keV were discarded to reduce the contamination
from the instrumental background due to Al ${\rm K_{\alpha}}$ and Si ${\rm K_{\alpha}}$ fluorescent lines \citep{Kuntz2008,Snowden2014}.
In both panels, the NXB was subtracted
and the X-ray background was subtracted with vignetting taken into account.
The distance is measured from the position of the pulsar.
The count rate of each bin was normalized (with vignetting taken into account) to that in the 
entire rectangle of $5' \times 1'$ closest to the pulsar in the PWN-west.
The count rate profiles obtained by \textit{Suzaku} (Figure~3) are also shown by open circles (shifted by $0.\hspace{-2pt}'1$),
after adjusting the normalization for comparison.
\label{fig:f1}
}
\end{figure}

\clearpage

\begin{figure}[ht!]
\figurenum{8}
\gridline{
\fig{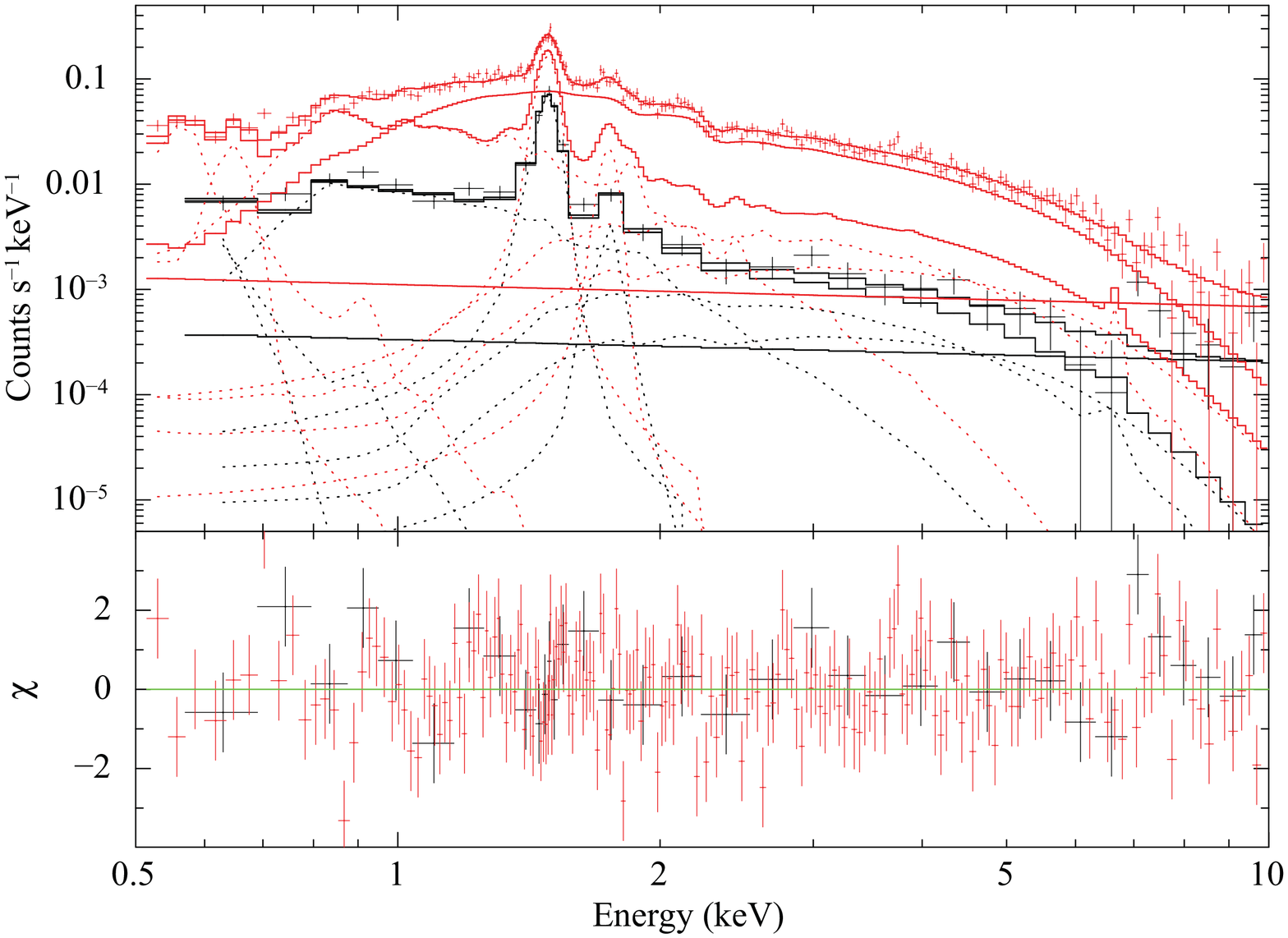}
{0.45\textwidth}{(a)}
}
\gridline{
\fig{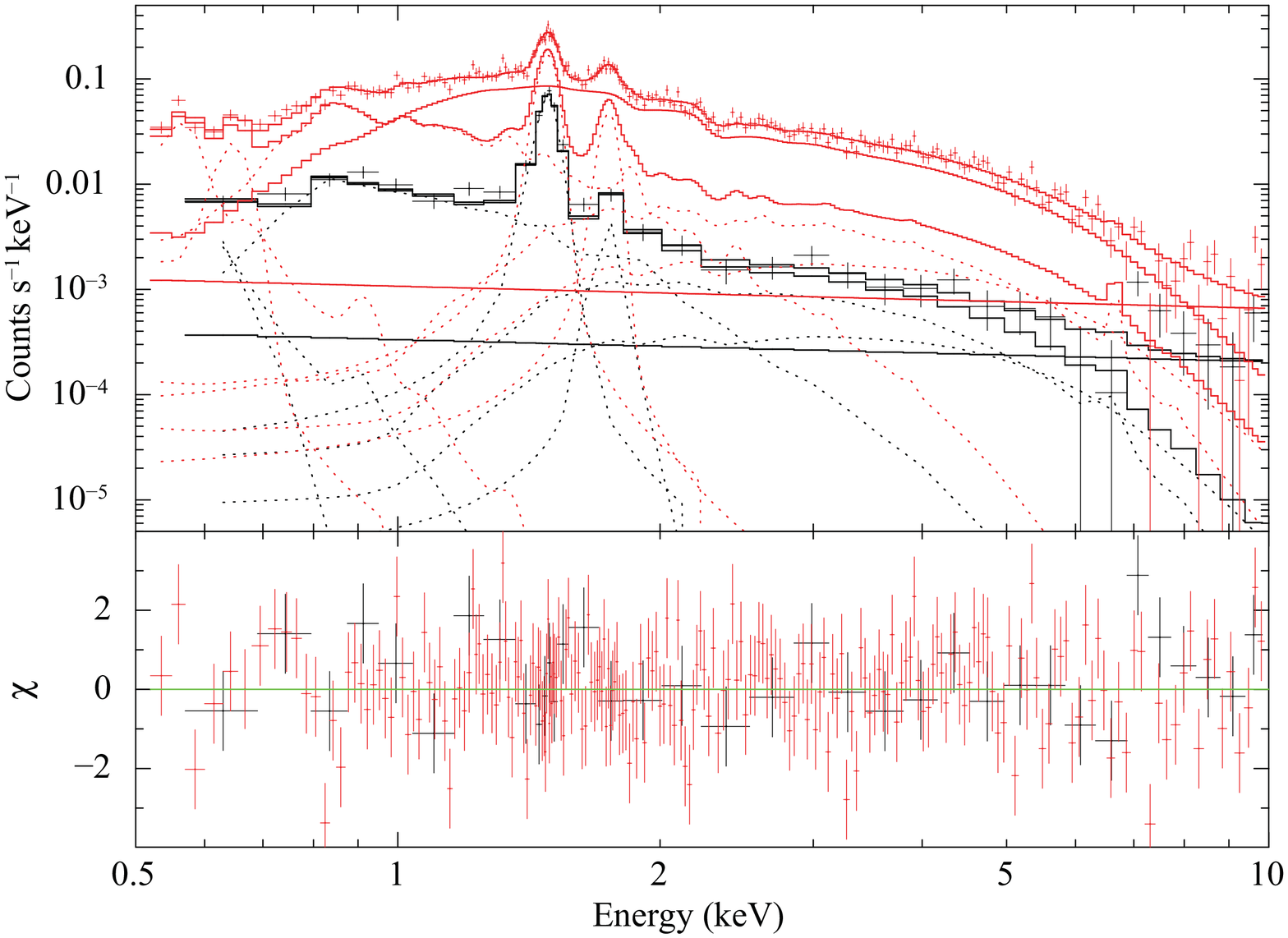}
{0.45\textwidth}{(b)}
}
\gridline{
\fig{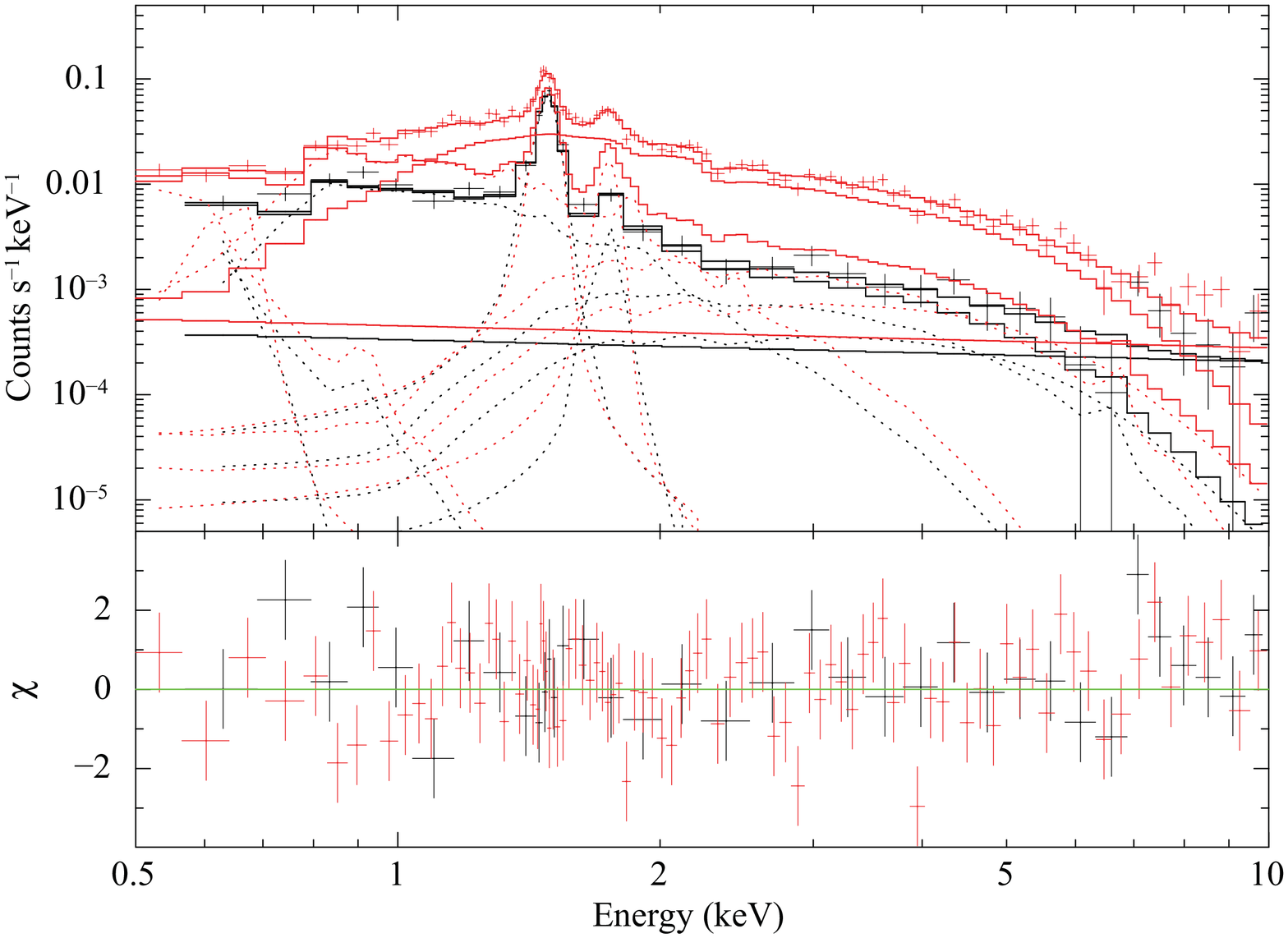}
{0.45\textwidth}{(c)}
}
\caption{
The PWN-west (panel (a)) and PWN-east (panel (b)) spectra of \textit{XMM-Newton} MOS, each integrated over $12' \times 5'$ (see Figure~6).
Also shown in panel (c) is the spectrum of the Arc integrated over $9' \times 2.\hspace{-2pt}'5$.
Red and black crosses are spectra of the source region and the background region, respectively,
and best-fit models (total) are indicated by red and black lines along the data points.
The solid lines in horizontal direction indicate estimated contributions of the residual soft protons.
The solid lines between the total model and the residual soft-proton contamination 
correspond to the PWN emission (in the source region) and the X-ray background (in the source and background regions)
and are further decomposed into each component (dotted lines).
\label{fig:f1}
}
\end{figure}

\clearpage

\begin{table}[htbp]
\caption{Summary of the spectral fits of the PWN-west, PWN-east, and Arc by \textit{XMM-Newton} MOS}
\begin{center}
\small
\begin{tabular}{cccccc} 
\hline\hline
Region & $N({\rm H})$ & $\Gamma$ & $f({\rm 0.5\mbox{--}2~keV})$ & $f({\rm 2\mbox{--}10~keV})$ & $\chi^{2}$/DOF \\
       & $10^{21}~{\rm cm^{-2}}$ &          & $10^{-13}~{\rm erg~s^{-1}~cm^{-2}}$ & $10^{-13}~{\rm erg~s^{-1}~cm^{-2}}$& \\ 
\hline
west    & $8.1\pm1.0$ & $2.10\pm0.12$ & $4.77^{+0.22}_{-0.30}$ & $19.3^{+0.9}_{-1.5}$ & 271.8/225 \\
east    & $7.5\pm0.9$ & $2.03\pm0.10$ & $5.01^{+0.25}_{-0.26}$ & $20.4^{+1.1}_{-1.3}$ & 304.4/244 \\
arc     & $9.3^{+1.5}_{-1.4}$ & $2.13\pm0.18$ & $1.66^{+0.12}_{-0.13}$ & $7.31^{+0.67}_{-0.79}$ & 138.8/116 \\ 
\hline
\end{tabular}
\tablenotetext{}{{\bf Notes:} The $N({\rm H})$ and $\Gamma$ are the hydrogen column density
of the photoelectric absorption and the photon index of the power-law model, respectively.
$f({\rm 0.5\mbox{--}2~keV})$ and $f({\rm 2\mbox{--}10~keV})$ are absorption-uncorrected fluxes in
0.5--2 and 2--10~keV, respectively.
Errors are calculated for single-parameter 90\% confidence limit.}
\end{center}
\end{table}

\begin{table}[htbp]
\caption{Summary of the spectral fits of the PWN-west and PWN-east divided into segments}
\begin{center}
\small
\begin{tabular}{ccccccc} 
\hline\hline
\multicolumn{2}{c}{Region} & $N({\rm H})$ & $\Gamma$ & $f({\rm 0.5\mbox{--}2~keV})$ & $f({\rm 2\mbox{--}10~keV})$ & $\chi^{2}$/DOF \\
                         & & $10^{21}~{\rm cm^{-2}}$ &          & $10^{-13}~{\rm erg~s^{-1}~cm^{-2}}$ & $10^{-13}~{\rm erg~s^{-1}~cm^{-2}}$& \\ 
\hline
west & $0'\mbox{--}3'$  & $7.8\pm1.1$ & $1.99^{+0.13}_{-0.14}$ & $1.71\pm0.09$ & $7.60^{+0.50}_{-0.53}$ & 123.0/104 \\
     & $3'\mbox{--}6'$  & $8.1^{+1.5}_{-1.3}$ & $2.07^{+0.19}_{-0.18}$ & $1.33\pm0.10$ & $5.57^{+0.47}_{-0.52}$ & 82.6/80 \\
     & $6'\mbox{--}9'$  & $8.3^{+1.7}_{-2.0}$ & $2.44^{+0.27}_{-0.33}$ & $1.23^{+0.09}_{-0.17}$ & $3.52^{+0.65}_{-0.53}$ & 110.5/73 \\
     & $9'\mbox{--}12'$ & $10.8^{+4.7}_{-3.9}$ & $2.53^{+0.73}_{-0.59}$ & $0.61^{+0.10}_{-0.20}$ & $2.00^{+0.41}_{-0.85}$ & 79.3/57 \\
     & $0'\mbox{--}3'$  & 8.1(fix) & $2.02\pm0.08$ & $1.70^{+0.10}_{-0.08}$ & $7.54^{+0.52}_{-0.43}$ & 123.3/105 \\
     & $3'\mbox{--}6'$  & 8.1(fix) & $2.07\pm0.10$ & $1.33^{+0.08}_{-0.07}$ & $5.58^{+0.48}_{-0.47}$ & 82.6/81 \\
     & $6'\mbox{--}9'$  & 8.1(fix) & $2.40\pm0.16$ & $1.24\pm0.10$ & $3.56^{+0.53}_{-0.46}$ & 110.4/74 \\
     & $9'\mbox{--}12'$ & 8.1(fix) & $2.18\pm0.30$ & $0.61^{+0.12}_{-0.11}$ & $2.25^{+0.64}_{-0.58}$ & 80.5/58 \\
\hline
east & $0'\mbox{--}3'$  & $8.3^{+1.4}_{-1.2}$ & $1.99\pm0.15$ & $1.30^{+0.08}_{-0.09}$ & $6.11^{+0.44}_{-0.40}$ & 109.1/96 \\
     & $3'\mbox{--}6'$  & $7.0^{+1.3}_{-1.2}$ & $2.11\pm0.17$ & $1.46\pm0.10$ & $5.21^{+0.51}_{-0.45}$ & 106.1/92 \\
     & $6'\mbox{--}9'$  & $7.8^{+1.8}_{-1.6}$ & $2.00^{+0.22}_{-0.21}$ & $1.08^{+0.07}_{-0.10}$ & $4.74^{+0.44}_{-0.54}$ & 75.2/74 \\
     & $9'\mbox{--}12'$ & $7.2^{+2.1}_{-1.8}$ & $1.97^{+0.27}_{-0.25}$ & $1.11^{+0.10}_{-0.14}$ & $4.75^{+0.76}_{-0.58}$ & 75.5/72 \\
     & $0'\mbox{--}3'$  & 7.5(fix) & $1.91^{+0.08}_{-0.09}$ & $1.31\pm0.09$ & $6.22^{+0.47}_{-0.41}$ & 110.4/97 \\
     & $3'\mbox{--}6'$  & 7.5(fix) & $2.17\pm0.10$ & $1.46\pm0.10$ & $5.13^{+0.39}_{-0.40}$ & 106.6/93 \\
     & $6'\mbox{--}9'$  & 7.5(fix) & $1.97\pm0.12$ & $1.08^{+0.09}_{-0.08}$ & $4.78^{+0.53}_{-0.40}$ & 75.3/75 \\
     & $9'\mbox{--}12'$ & 7.5(fix) & $2.00\pm0.15$ & $1.11^{+0.10}_{-0.11}$ & $4.71^{+0.51}_{-0.60}$ & 75.6/73 \\ 
\hline
\end{tabular}
\tablenotetext{}{{\bf Notes:} See Notes of Table~3 for the detailed description of the parameters.}
\end{center}
\end{table}

\clearpage

\subsection{Summary of X-ray Data Analysis Results}
Before proceeding to the discussion (Section~4), let us summarize the results of the X-ray data analysis.
\begin{enumerate}


\item Even with \textit{Suzaku}-XIS, no extended emission was found in the western region
of TeV emission
(Section~3.1.1).

\item The source extent of the PWN-west was measured to be $15' \times 10'$ by \textit{Suzaku}-XIS,
with a linear decrease of the intensity from $0'$ to $15'$ (Section~3.1.1).
The \textit{XMM-Newton} data indicate that the PWN-east has a flatter intensity profile up to $12'$, beyond which the
source extent is not constrained. The Arc has a source extent of ${\sim}9' \times 2.\hspace{-2pt}'5$. (Section~3.2.1)

\item The orientation of the PWN major axis is ${\sim}71\fdg4$ east from the north (Section~3.1.1).

\item The PWN-west spectrum is represented by an absorbed power-law with 
$N({\rm H}) = (8.2^{+1.3}_{-1.1}) \times 10^{21}~{\rm cm^{-2}}$ and $\Gamma = 2.05\pm0.12$.
No significant change of $\Gamma$ was found inside the region (Section~3.1.2). 
With the results of \textit{XMM-Newton} spectral analysis (Section~3.2), 
we confirm that the PWN-east and the Arc have similar spectral parameters to those of the PWN-west. (Section~3.2.2)

\item The 2--10~keV observed flux $f({\rm 2\mbox{--}10~keV})$
in the region of $15' \times 5'$ to the west of the PWN was measured to be  
$2.6 \times 10^{-12}~{\rm erg~s^{-1}~cm^{-2}}$ (Section~3.1.2),
giving the absorption-corrected flux
$F({\rm 2\mbox{--}10~keV})$ of $2.8 \times 10^{-12}~{\rm erg~s^{-1}~cm^{-2}}$.
On the basis of the morphology along the minor axis
(Section~3.1.1), we obtained $F({\rm 2\mbox{--}10~keV}) \sim 4.1 \times 10^{-12}~{\rm erg~s^{-1}~cm^{-2}}$
for the entire PWN-west.
Although it was not possible to constrain the extent of the PWN-east beyond the \textit{XMM-Newton} FOV, the
results of \textit{XMM-Newton} spectral analysis (Section~3.2.2) suggest that
the PWN-east plus Arc has $F({\rm 2\mbox{--}10~keV}) \sim 3.0 \times 10^{-12}~{\rm erg~s^{-1}~cm^{-2}}$ at least,
giving the lower limit of $F({\rm 2\mbox{--}10~keV})$ for
the overall PWN (the PWN-west, PWN-east, and Arc) to be
${\sim}7 \times 10^{-12}~{\rm erg~s^{-1}~cm^{-2}}$.

\end{enumerate}

\clearpage

\section{Discussion}

\subsection{Properties of the X-ray PWN}
Here we describe the properties of the X-ray PWN, and the implications of these properties,
without discussing its
relation with VER~J2019+368.

The photon index of the PWN-west we obtained, $\Gamma=2.05\pm0.12$, is significantly larger than
that of the PWN emission close to the pulsar measured by \textit{Chandra}
as reported by \citet{Etten2008}; they obtained $\Gamma=1.0\mbox{--}1.5$ within ${\sim}10''$ from the pulsar (``Inner nebula''),
and $\Gamma \sim 1.7$ for their ``Jet'' and ``Outer nebula-east'' (which is within ${\sim}1.5'$ of the pulsar).
The photon index we measured is consistent, however, with $\Gamma = 1.93\pm0.13$ for their 
``Outer nebula-west'' (which is 1.5--3$'$ from the pulsar).
As no significant change of $\Gamma$ is observed up to $15'$ from the pulsar toward the southwest direction in \textit{Suzaku} data,
we can conclude that the CR electrons accelerated at the PWN termination shock
\citep[${\sim}10''$ from the pulsar; ][]{Etten2008}
suffer from synchrotron cooling close to the pulsar (within ${\sim}1.5'$)
but propagate outward without significant cooling.
The \textit{XMM-Newton} data indicate similar conclusions on the PWN-east; the photon index ($\Gamma = 2.03 \pm 0.10$)
is larger than that of the pulsar, Jet, and Outer nebula-east and no significant change of $\Gamma$ is observed
up to $12'$ toward the northeast.

Through observations by \textit{Suzaku} (Section~3.1),
\textit{XMM-Newton} (Section~3.2), and \textit{Chandra} \citep{Etten2008},
the absorption of the X-ray PWN was found to be $(6\mbox{--}9) \times 10^{21}~{\rm cm^{-2}}$,
significantly lower than the total Galactic absorption 
in the direction of Cygnus-X of ${\ge}2.5 \times 10^{22}~{\rm cm^{-2}}$ estimated by
\citet{Mizuno2015} using X-ray source spectra and $\gamma$-ray data.
Therefore, the pulsar and its PWN are unlikely to be located at a distance ${\ge}10~{\rm kpc}$,
as was inferred from radio data (see Section~1). 
Instead, we adopt the distance $d = 1.8^{+1.7}_{-1.4}~{\rm kpc}$ estimated by \citet{Kirichenko2015}
based on the absorption-distance relation using red-clump stars in the direction of the pulsar.

\clearpage

\subsection{Relation to VER~J2019+368}
We first discuss implications of the determined properties in the
X-ray and TeV $\gamma$-ray regimes.
We then examine particle transport (and magnetic fields), primarily within the X-ray PWN, and implications.
We finally present a possible model to explain the multiwavelength data.

First, the fact that the major axes of the X-ray PWN and VER~J2019+368 are almost parallel (Section~3.1.1)
strongly support that the X-ray PWN is physically associated with VER~J2019+368.
If the X-ray PWN is a counterpart of the TeV emission, 
TeV $\gamma$ rays are likely to be produced by the inverse Compton (IC) scattering by X-ray producing CR electrons.
In the case of the PWN synchrotron/IC scenario,
temporarily neglecting the details of the electron spectrum and the Klein-Nishina effect
produces a ratio of X-ray to TeV $\gamma$-ray luminosities given by the ratio of magnetic field
energy density to photon field energy density, $U_{\rm mag}/U_{\rm ph}$.
The absorption-corrected PWN flux in the X-ray regime, $F({\rm 2\mbox{--}10~keV}) \sim 7 \times 10^{-12}~{\rm erg~s^{-1}~cm^{-2}}$,
is close to the TeV $\gamma$-ray flux of VER~J2019+368 
[$F({\rm 1\mbox{--}10~TeV}) \sim 6.7 \times 10^{-12}~{\rm erg~s^{-1}~cm^{-2}}$],
indicating either that $U_{\rm mag}$ is close to the energy density of the cosmic microwave background (CMB)
or that the average magnetic field $B$ of the PWN is rather low and close to the typical interstellar magnetic field,
$B \sim 3~\mu{\rm G}$.
This value should be taken as a lower limit since the X-ray observations did not cover the whole
TeV-emitting region (see Figure~1). Nevertheless, a much larger value of the magnetic field is
unlikely, since the two \textit{Suzaku} observations covered the central region of the TeV emission.
Hereafter we express physical quantities with $B$ normalized by $3~\mu{\rm G}$.

From the spectra and morphologies in the X-ray and TeV $\gamma$-ray regimes, we can constrain the properties
of the accelerated CR electrons. Hereafter, we assume a constant injection of accelerated
CR electrons into uniform magnetic and radiation fields over the lifetime of the pulsar for simplicity.

From discussions in, e.g., \citet[][]{Longair2011} and \citet[][]{Jager2008},
the characteristic energy of X-rays owing to synchrotron radiation 
($\epsilon_{\rm X}$) in the magnetic field $B$ is related to the electron energy, $E_{e}$, as
\begin{linenomath*}
\begin{equation}
E_{e} \simeq 132~{\rm TeV}~\left( \frac{\epsilon_{\rm X}}{1~{\rm keV}} \right)^{0.5}~\left( \frac{B}{3~\mu{\rm G}} \right)^{-0.5}~~.
\end{equation}
\end{linenomath*}
On the other hand, the typical energy of $\gamma$ rays ($\epsilon_{\gamma}$)
generated by IC scattering of the CMB photons is related to $E_{e}$ as,
\begin{linenomath*}
\begin{equation}
E_{e} \simeq 17.2~{\rm TeV}~\left( \frac{\epsilon_{\gamma}}{1~{\rm TeV}} \right)^{0.5}~~.
\end{equation}
\end{linenomath*}
Equations~(1) and (2) indicate that electrons with $E_{e} \ge 100~{\rm TeV}$ are required to generate synchrotron X-rays
above 1~keV with $B \sim 3~\mu{\rm G}$,
while electrons with $E_{e} \le 50~{\rm TeV}$ produce $\gamma$ rays below 10~TeV.
As we obtained $\Gamma=2.05$ for the X-ray PWN and $\Gamma=1.75$ was reported for TeV $\gamma$-ray emission,
there must be a spectral break of accelerated CR electrons at around 50--100~TeV.
Although the TeV $\gamma$-ray photon index has a rather large uncertainty of 
${\sim}0.2$ \citep{Aliu2014}, the Klein-Nishina effect softens the electromagnetic spectrum 
more than in the Thomson regime,
in which the photon index of IC emission is the same as that of the synchrotron radiation 
with the same CR electron spectral index.
Therefore, our hypothesis of a spectral break is robust.
The process causing this spectral break is likely to be a synchrotron cooling, in which 
the break energy $E_{\rm bk}$ is related to the injection time $t_{0}$ and magnetic field $B$ as
\begin{linenomath*}
\begin{equation}
E_{\rm bk} \simeq 80~{\rm TeV} \left( \frac{t_{0}}{17.2~{\rm kyr}} \right)^{-1} \left( \frac{B}{3~\mu{\rm G}} \right)^{-2}~~,
\end{equation}
\end{linenomath*}
where $t_{0}$ is normalized to the characteristic age of the pulsar.
The electron spectral index changes by 1, and the synchrotron radiation
and IC emission each change by 0.5,
indicating that 
the difference in spectral slopes between X-rays and TeV $\gamma$ rays can be naturally explained
by the canonical age of the pulsar (characteristic age of 17.2~kyr) and $B=3~\mu{\rm G}$.
We should also take into account the cooling of CR electrons during propagation;
for electrons producing X-rays above 1~keV, the main mechanism for this is synchrotron cooling.
Then, using Equation~(1), the cooling time $\tau$ can be expressed as
\begin{linenomath*}
\begin{equation}
\tau(\epsilon_{\rm X}) \simeq 10.5~{\rm kyr} \left( \frac{\epsilon_{\rm X}}{1~{\rm keV}} \right)^{-0.5}
\left( \frac{B}{3~\mu{\rm G}} \right)^{-1.5}~~.
\end{equation}
\end{linenomath*}
This indicates that the lifetimes of CR electrons producing X-rays at 1 and 10~keV
(under $B=3~\mu{\rm G}$) are 10.5 and 3.3~kyr, respectively. If we also take into account the cooling
owing to IC scattering of the CMB and infrared background
(based on the blackbody radiation at a temperature of 30~K and energy density of $0.3~{\rm eV~cm^{-3}}$; see below)
using the procedure described in \citet[][]{Moderski2005}, the true lifetimes of electrons
producing 1 and 10~keV X-rays are found to be 7.9 and 3.0~kyr, respectively. 
Therefore, Equation~(4) is valid to within 25\%.

Let us then discuss particle transport and its implications. 
The CR electrons are transported via either diffusion or advection caused by the pulsar wind.
If advection is the dominant process, high-energy electrons with shorter lifetimes [Equation~(4)]
will make it closer from the pulsar.
This implies a spectral softening not seen in our detailed study of the X-ray spectrum
(Section~3.1.2).
Therefore,
the highest energy electrons we consider propagate over a distance ${\ge}15'$ during their lifetime.
Since the angular extent of $15'$ corresponds to $8~{\rm pc} \left( \frac{d}{1.8~{\rm kpc}}\right)$,
and the lifetime of electrons producing 10~keV X-rays due to synchrotron radiation is 
$3.3~{\rm kyr} \left( \frac{B}{3~\mu{\rm G}}\right)^{-1.5}$,
the advection velocity divided by the speed of light ($\beta_{\rm adv}$) should satisfy
\begin{linenomath*}
\begin{equation}
\beta_{\rm adv} \ge 7.9 \times 10^{-3} 
\left( \frac{B}{3~\mu{\rm G}} \right)^{1.5}
\left( \frac{d}{1.8~{\rm kpc}} \right)~~.
\end{equation}
\end{linenomath*}
In this scenario, the absence of
X-ray emission beyond the peak position of the TeV emission is due to the lower surface brightness
of synchrotron X-rays caused by the lower magnetic field or lower CR electron density.
The scenario can naturally explain the
larger size of TeV emission produced by electrons of lower energy (longer lifetime).
In the case of diffusion-dominated scenario, the electrons propagate the diffusion length
of $\sqrt{2D\tau}$, where $D$ and $\tau$ are the diffusion coefficient and the electron lifetime, respectively.
Then we can constrain $D$ as we did to constrain $\beta_{\rm adv}$.
Let us first examine the case of energy-independent diffusion,
as predicted by, e.g., \citet[][]{Porth2016}
through three-dimensional magnetohydrodynamics simulations.
In order for the diffusion length 
to exceed the length of the X-ray PWN,
even for electrons producing 10~keV X-rays, we obtain
\begin{linenomath*}
\begin{equation}
D \ge 2.9 \times 10^{27}~{\rm cm^{2}~s^{-1}} 
\left( \frac{B}{3~\mu{\rm G}} \right)^{1.5}
\left( \frac{d}{1.8~{\rm kpc}} \right)^{2}~~.
\end{equation}
\end{linenomath*}
Like the advection-dominated scenario,
the absence of X-ray emission beyond the TeV emission peak is due to the lower magnetic field or lower CR electron density,
and the larger size of TeV emission is due to the cooling of electrons producing X-rays.
Alternatively, diffusion can naturally explain the apparent lack of spectral softening,
if $D$ depends on the particle energy as
$D \propto E_{\rm e}^{\delta}$ with $\delta \sim 1$ \citep[e.g.,][]{Etten2011}.
If $\delta=1$
the diffusion coefficient $D$ can be expressed as
$D = \frac{1}{3}\lambda_{g}c\eta$, where $\lambda_{g}$ is the electron gyroradius,
$c$ is the speed of light, and the parameter $\eta$ is related to
the degree of magnetic turbulence. 
By substituting the physical constants and also using Equation~(1), we obtain
\begin{linenomath*}
\begin{eqnarray}
D & = &
1.11 \times 10^{27} \eta ~{\rm cm^{2}~s^{-1}}
\left(\frac{E_{\rm e}}{100~{\rm TeV}}\right)
\left( \frac{B}{3~\mu{\rm G}} \right)^{-1} \nonumber \\
 & \simeq &
1.46 \times 10^{27} \eta~{\rm cm^{2}~s^{-1}} \left( \frac{\epsilon_{\rm X}}{1~{\rm keV}}\right)^{0.5}
\left( \frac{B}{3~\mu{\rm G}} \right)^{-1.5}~~.
\end{eqnarray}
\end{linenomath*}
Then, in order for the source extent not to exceed 
the diffusion length in electron lifetime ($\sqrt{2D\tau}$), we obtain 
[by substituting Equation~(4)]
\begin{linenomath*}
\begin{equation}
\eta \simeq 0.60 \left( \frac{B}{3~\mu{\rm G}}\right)^{3} \left( \frac{d}{1.8~{\rm kpc}} \right)^{2}~~.
\end{equation}
\end{linenomath*}
Therefore, under the condition of $B \sim 3~\mu{\rm G}$ and $d \sim 1.8~{\rm kpc}$,
$\eta \sim 1$ (i.e., close to Bohm limit) is required,
suggesting that the magnetic field is highly turbulent.
Energy-dependent diffusion alone, however, is not able to explain the larger size of the TeV emission.
We thus constrain the advection velocity $\beta_{\rm adv}$ or the diffusion coefficient $D$ 
from the morphology of the X-ray PWN.

On the basis of the discussions above (in particular, regarding $E_{\rm bk}$ and $B$), 
under the assumption of the constant injection and uniform magnetic field,
we present a possible multiwavelength spectral model in Figure~9
in which the electron spectrum is assumed to be a power-law with a photon index of 2.1 below 80~TeV and 3.1 above 80~TeV
and with an exponential cutoff at 1~PeV.
Contributions from synchrotron radiation and IC scattering are computed based on
\citet{SyncModel} and \citet{ICModel} respectively.
We adopted $B=3~\mu{\rm G}$ and adjusted the model normalization to explain the entire X-ray PWN flux 
in the region 2--10~keV
(${\sim}7 \times 10^{-12}~{\rm erg~s^{-1}~cm^{-2}}$; see Section~3.3).
To calculate the IC scattering emission, we referred to the radiation field model of \citet{Porter2008}.
Because the source distance from the Galactic center is estimated to be 8.2~kpc
in our case where $l=75\arcdeg$ and $d=1.8~{\rm kpc}$, we adopted their model 
at the solar circle and assumed a CMB and
infrared background with temperature 30~K and energy density $0.3~{\rm eV~cm^{-3}}$.
We also overlaid the $\gamma$-ray emission from the pulsar and an upper limit of the PWN in the GeV band
taken from \citet{Abdo2009}.
\citet{Paredes2009} reported extended radio emission of ${\sim}700~{\rm mJy}$ at 1.4~GHz in the vicinity of VER~J2019+368.
Although they did not provide information on the position and spatial extent, we plot their flux for reference.
As is seen from Figure~9, the model explains about 80\% of the TeV emission,
indicating that the X-ray PWN is a major contributor to VER~J2019+368.
Considering the assumptions we made (constant injection of CR electrons into uniform magnetic and radiation fields
over the pulsar lifetime),
limited coverage of X-ray observations (see Figure~1), 
and the apparent offset of the pulsar from the peak of the TeV emission
(which cannot be explained by our simplified scenario),
we do not rule out X-ray emission from the nebula further out and/or
the confusion of TeV source(s) physically unrelated to the X-ray PWN.
Further observations in X-rays and TeV $\gamma$ ray are worthwhile to fully understand the system.
In particular, TeV $\gamma$-ray observations at the better sensitivity and angular resolution
by the Cherenkov Telescope Array (CTA) \citep{Actis2011} are anticipated 
to reveal the TeV $\gamma$-ray properties in more detail.

In the discussions above, we have assumed constant injection of electrons into uniform magnetic fields for simplicity.
If the pulsar has already experienced significent energy losses, it injected more electrons in the past,
therefore the ratio of TeV to X-ray flux is increased. In order for the predicted TeV flux not to exceed the observed value,
a magnetic field larger than $B=3~\mu{\rm G}$ is required. In this case, $E_{\rm bk} \sim 80~{\rm TeV}$ 
(which is required to explain the different spectral indices between X-ray and $\gamma$ ray) can be achieved
if the true age of the pulsar is younger (see also Section~4.3).

So far we have assumed only the PWN since no evidence of a host SNR is found \citep{Etten2008}.
If the parent SNR is found in future, the discussion on particle transport and the relation between
X-ray and TeV $\gamma$ ray might be affected.

\clearpage

\begin{figure}[ht!]
\figurenum{9}
\centering
\includegraphics[width=0.6\textwidth]{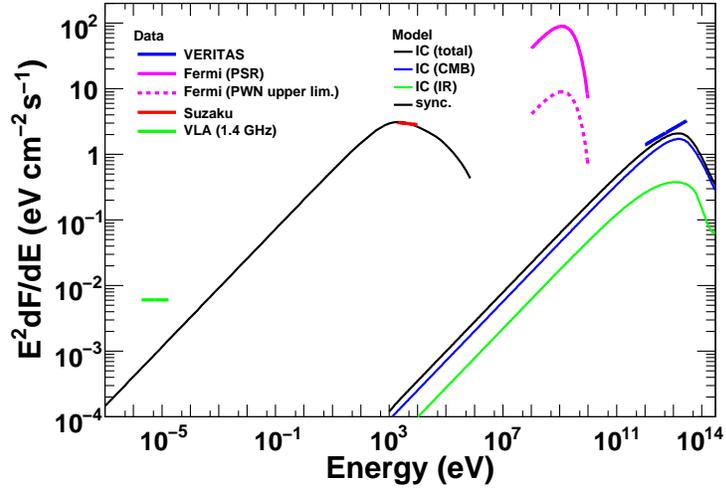}
\caption{
Multiwavelength spectrum and model.
Data are from X-ray (\textit{Suzaku} and \textit{XMM-Newton}), TeV $\gamma$ rays, GeV $\gamma$ rays and radio measurements.
Model assumes $B=3~\mu{\rm G}$ and a spectral break at 80~TeV from $\Gamma=2.1$ to 3.1. 
The thin blue and green lines show contributions of IC scattering off the CMB and infrared background, respectively.
See text for details.
\label{fig:f1}
}
\end{figure}

\clearpage

\subsection{Comparison with Other PWNe}

Finally, we compare the properties of the X-ray PWN and VER~J2019+368 with other X-ray PWNe associated with TeV $\gamma$ rays.
According to \citet{Mattana2009} who compiled the properties of 14 PWNe,
the $\gamma$ ray to X-ray energy flux ratio is approximately
proportional to the pulsar characteristic age, owing to the effect of severe cooling 
on X-ray production.
The energy flux ratio at 1--30~TeV and 2--10~keV in our case is
$(11.6 \times 10^{-12})/(7 \times 10^{-12}) \sim 1.6$, which is roughly consistent with their Figure~1.
Therefore, the smaller flux and size of the X-ray region can be understood,
as with other TeV-emitting PWNs,
to be caused by faster cooling of X-ray-producing electrons.
\citet{Bamba2010} studied eight PWNs with various characteristic ages,
which are associated with TeV $\gamma$-ray sources;
in particular, they studied the size of an X-ray PWN as a function of the pulsar characteristic age.
They found a rather constant size up to $10~{\rm kyr}$ and then a gradual increase 
in size thereafter, 
possibly caused by an increase in advection speed or a decrease in the magnetic field
turbulence as the pulsar/PWN grows older.
Our measured size of the X-ray PWN, $8~{\rm pc} \left( \frac{d}{1.8~{\rm kpc}}\right)$,
is consistent with those of their samples of similar characteristic age.

We should also compare with the archetype evolved PWN HESS~J1825-137 and its extended X-ray PWN around
PSR~J1826-1334. The properties of the pulsar are similar to those of PSR~J2021+3651.
The pulse period and its derivatives are $101~{\rm ms}$ and $7.5 \times 10^{-14}$, respectively
\citep{Clifton1992}, giving the surface magnetic field of $2.8 \times 10^{12}~{\rm G}$,
the characteristic age of 21.4~kyr, and
spin-down luminosity of $2.8 \times 10^{36}~{\rm erg~s^{-1}}$.
The PWN has also similar properties.
The TeV PWN extends more than $1\arcdeg$ with the $\gamma$-ray peak position
being offset from the pulsar and X-ray peak position by ${\sim}10'$ \citep{Aharonian2006}.
The source extent of the X-ray PWN was measured by \textit{Suzaku} \citep{Uchiyama2009,Etten2011}
to be ${\sim}15'$ towards the south \citep[the northern part of the pulsar has not been observed; see Figure~1 of ][]{Etten2011}.
While the compact core of the X-ray PWN has a hard photon index of ${\sim}1.6$ \citep{Gaensler2003},
the outer part of the X-ray PWN does not exhibit significant spectral softening with 
the photon index of ${\sim}2$ \citep{Uchiyama2009}.
There are, however, two distinct properties between two systems.
Firstly, the energy fluxes in X-ray (2--10~keV) and TeV $\gamma$ ray (1--10~TeV) of HESS~J1825-137 are
$4.5 \times 10^{-12}$ and $51 \times 10^{-12}~{\rm erg~s^{-1}~cm^{-2}}$, respectively \citep{Uchiyama2009,Aharonian2006},
giving ${\sim}10$ times larger $F(1\mbox{--}10~{\rm TeV})/F(2\mbox{--}10~{\rm keV})$ ratio
than that we obtained for VER~J2019+368.
Secondly, the photon index in TeV $\gamma$ ray of HESS~J1825-137 is ${\sim}2.4$ on average
and is significantly softer than that of VER~J2019+368.
These two facts can be explained naturally by assuming more severe cooling of electrons in HESS~J1825-137.
If the CR electrons of energies less than ${\sim}50~{\rm TeV}$ have already suffered from cooling,
a softer spectral index in TeV and a larger ratio of the TeV $\gamma$-ray flux to the X-ray flux are expected.
Probably VER~J2019+368 has weaker mean magnetic field and/or the true age of the pulsar is younger.
We also note that \citet{Etten2011} carried out a modeling of X-ray and TeV $\gamma$-ray data
of HESS~J1825-127 and constrained the electron injection history, profile of the magnetic field,
advection velocity, and diffusion coefficient. Although such an extensive modeling is beyond the scope of our study,
detailed study of the morphology in TeV $\gamma$ ray by future observations by CTA is anticipated to better
understand the VER~J2019+368 system.

\clearpage

\section{Summary}

We conducted deep X-ray observations of the VER~J2019+368 region using
\textit{Suzaku}-XIS to examine the properties of the X-ray PWN around PSR~J2021+3651
and to search for previously unknown extended X-ray emissions.
We also analyzed archival \textit{XMM-Newton} data
to complement the \textit{Suzaku} observations, which did not cover the entire region of VER~J2019+368.
We found that the total size of the X-ray PWN along the major axis is more than $27'$:
the PWN-west have a source extent of approximately $15' \times 10'$ with an orientation of its major axis
nearly parallel to that of TeV emission, and the PWN-east extends up to at least $12'$ from the pulsar.
The PWN spectra were well fitted by an absorbed power-law
for absorption at ${\sim}8 \times 10^{21}~{\rm cm^{-2}}$ and a photon index of ${\sim}2$,
with no obvious change in the index occurring within the X-ray PWN.
The measured X-ray absorption favors the distance to the source to be
much smaller than $10~{\rm kpc}$ inferred from radio data.
Aside from the PWN around PSR~J2021+3651, no extended emission was found by even \textit{Suzaku}-XIS. 
The uniformity of the X-ray photon index constrains the advection velocity or the diffusion coefficient
depending on the primary process of
particle transport for X-ray-producing CR electrons.
From the measured X-ray spectrum, reported TeV $\gamma$-ray spectrum and X-ray source extent,
under the assumption of the constant injection of CR electrons into the uniform magnetic and radiation fields
over the characteristic age of the pulsar,
we obtained a rather low magnetic field of ${\sim}3~\mu{\rm G}$.
Our synchrotron/IC model is able to explain ${\sim}80\%$ of the TeV emission, 
indicating that the X-ray PWN is a major contributor to VER~J2019+368. 
To fully understand the nature of the extended TeV emission, higher-sensitivity and 
higher-resolution observations
by facilities such as CTA will be useful.

We would like to thank S. Kisaka for valuable comments, and H. Katagiri 
for helping the calculation of multiwavelength spectral model.
We also thank the referee for his/her valuable comments, and the \textit{Suzaku} team members and \textit{XMM-Newton} team members
for their dedicated support of the satellite operation and calibration.
This work was partially supported by JSPS Grant-in-Aid for Scientific Research Grant Numbers 
JP25287059 (T.M.), JP15K05088 (R.Y.), and JP26800160 (K. H.).

\clearpage

\appendix

\section{Detailed Descriptions and Parameters of the \textit{XMM-Newton} Data Analysis}
In order to estimate the residual soft-proton background (Section~3.2), we accumulated the spectrum
from the entire FOV and modeled it with a model which consists of a simple power-law 
({\tt pow} in {\tt XSPEC} to model the residual soft-proton contamination),
an absorbed power-law (${\tt wabs} \times {\tt pow}$ to model the CXB), 
thin-thermal plasma emission (${\tt wabs} \times {\tt apec}$ to model the hard-temperature emission of the GRXE),
two absorbed power-laws (${\tt wabs} \times {\tt pow}$ to reproduce the spectra of 
inner nebula and outer nebula reported by \citet{Etten2008}),
and another absorbed power-law (${\tt wabs} \times {\tt pow}$ to approximate the sum of 
point sources, emission from the pulsar, and rest of the PWN emission). 
The response matrix of diagonal unity elements is assumed in the first component.
For simplicity, a flat intensity profile is assumed
for the others. Some parameters were fixed to typical values 
(parameter values of the CXB were taken from \citet{Kuntz2008},
those of the GRXE were referred to \citet{Mizuno2015},
and those of the inner/outer nebulae were taken from \citet{Etten2008}).
Since we aim to constrain the residual 
soft-proton background which is prominent in high energy, we focused on data in 3--12~keV.
The obtained best-fit parameters are summarized in Table~A1. In the rest of the \textit{XMM-Newton} data analysis,
the spectral index of the residual soft-proton background is fixed to what obtained here 
($\Gamma = 0.21$)
with the normalization scaled using the {\tt proton-scale} command.

We then estimated the X-ray background to examine the count rate profile
by calculating the count rate of the background region ($5' \times 4'$ rectangle located in the northwest of the pulsar shown in Figure~6)
after the QPB and the residual soft-proton background (estimated from the entire FOV as above) was subtracted.
The background count rate was then subtracted from the count rate in the source region with the vignetting
taken into account. The obtained count rate profiles of the PWN are summarized in Figure~7,
in which data in 1.4--1.6 and 1.7--1.8~keV were discarded to reduce the contamination
from the instrumental background due to Al ${\rm K_{\alpha}}$ and Si ${\rm K_{\alpha}}$ fluorescent lines \citep{Kuntz2008,Snowden2014}.

We finally analyzed the spectrum of the PWN-west, PWN-east, and Arc. 
In addition to the CXB (${\tt wabs} \times {\tt pow}$) and high-temperature emission of the GRXE (${\tt wabs} \times {\tt apec}$),
we added two more thin-thermal plasma models (${\tt wabs} \times {\tt apec}$
to model the soft-temperature emission of the GRXE and local diffuse X-ray emission),
gaussian lines ({\tt gauss}) at 1.49 and 1.75~keV to model the fluorescent background
of Al and Si, and a line ({\tt gauss}) at 0.65~keV to model
the contribution of the Solar-Wind Charge eXchange \citep[SWCX,][]{Kuntz2008,Snowden2014}.
Those X-ray backgrounds were estimated by carrying out the joint spectral-fitting over the background region
and the source region, with the parameters coupled with the vignetting taken into account.
Again some parameters were fixed to typical values.
The obtained best-fit parameters of the background region are summarized in Table~A2, and those of the sources are given in Table~3.

\clearpage

\begin{table}[htbp]
\caption{Summary of the spectral fit of the entire FOV data}
\begin{center}
\small
\begin{tabular}{cc} 
\hline\hline
Parameter & Value \\ 
\hline
$N({\rm H})_{\rm CXB}\ (10^{21}~{\rm cm^{-2}})$ & 30(fixed) \\
$\Gamma_{\rm CXB}$ & 1.46(fixed) \\
${\rm Norm_{CXB}}$ & $5.79 \times 10^{-4}$(fixed) \\
$N({\rm H})_{\rm high} (10^{21}~{\rm cm^{-2}})$ & 30(fixed) \\
$kT_{\rm high}\ ({\rm keV})$ & 2.5(fixed) \\
$A_{\rm high}\ (Z_{\sun})$ & 0.3(fixed) \\
${\rm EM_{high}}$ & $2.5 \times 10^{-3}(\le 9.3 \times 10^{-3})$ \\
$N({\rm H})_{\rm in}\ (10^{21}~{\rm cm^{-2}})$ & 6.7(fixed) \\
$\Gamma_{\rm in}$ & 1.45(fixed) \\
${\rm Norm_{in}}$ & $6.0 \times 10^{-5}$(fixed) \\
$N({\rm H})_{\rm out}\ (10^{21}~{\rm cm^{-2}})$ & 6.7(fixed) \\
$\Gamma_{\rm out}$ & 1.82(fixed) \\
${\rm Norm_{out}}$ & $2.15 \times 10^{-4}$(fixed) \\
$N({\rm H})_{\rm src}\ (10^{21}~{\rm cm^{-2}})$ & $1.5(\le 40.7)$ \\
$\Gamma_{\rm src}$ & $1.84^{+0.87}_{-0.16}$ \\
${\rm Norm_{src}}$ & $(4.1^{+11.8}_{-1.7}) \times 10^{-3}$ \\
$\Gamma_{\rm SP}$ & $0.21^{+1.16}_{-0.51}$ \\
${\rm Norm_{SP}}$ & $(1.1^{+9.0}_{-1.0}) \times 10^{-2}$ \\ 
$\chi^{2}$/DOF & 174.1/142 \\ 
\hline
\end{tabular}
\tablenotetext{}{{\bf Notes:} 
$N({\rm H})_{\rm high}$, $kT_{\rm high}$, $A_{\rm high}$, and ${\rm EM_{high}}$ are
the absorption, temperature, abundance, and emission measure of the
high-temperature plasma models for the GRXE.
The emission measures are given as the value
integrated over the line of sight and the FOV, 
$\frac{\Omega}{4\pi} \int n_{\rm e} n_{\rm H} ds$ (where $n_{\rm e}$ and $n_{\rm H}$ are the electron
and hydrogen density, respectively, and $\Omega$ is the solid angle) in the unit of ${\rm 10^{14}~cm^{-5}}$.
$N({\rm H})_{\rm CXB}$, $\Gamma_{\rm CXB}$, and ${\rm Norm_{CXB}}$ are the absorption,
photon index of the power-law model, and intensity (${\rm photons~s^{-1}~cm^{-2}~keV^{-1}}$ at 1~keV)
integrated over the FOV 
for the CXB model, respectively. 
$N({\rm H})_{\rm in}$/$\Gamma_{\rm in}$/${\rm Norm_{in}}$ and 
$N({\rm H})_{\rm out}$/$\Gamma_{\rm out}$/${\rm Norm_{out}}$ are the absorption, photon index, and normalization
(${\rm photons~s^{-1}~cm^{-2}~keV^{-1}}$ at 1~keV)
of the inner and outer nebulae reported by \textit{Chandra} \citep{Etten2008}.
The same parameters with subscript src are for the additional absorbed power-law model 
to approximate the sum of point sources, emission from the pulsar, and rest of the PWN emission.
$\Gamma_{\rm SP}$ and ${\rm Norm_{SP}}$ are for the power-law model to represent the residual soft-proton background
convolved with the response matrix of diagonal unity elements.
Errors are calculated for single-parameter 90\% confidence limit.}
\end{center}
\end{table}

\clearpage
\begin{table}[htbp]
\caption{Summary of the spectral fits of the background region obtained through
a joint-fit with the PWN-west, PWN-east, and Arc}
\begin{center}
\small
\begin{tabular}{cccc} 
\hline\hline
 & PWN-west & PWN-east & Arc \\ 
\hline
$N({\rm H})_{\rm CXB} (10^{21}~{\rm cm^{-2}})$ & 30(fixed) & 30(fixed) & 30(fixed) \\ 
$\Gamma_{\rm CXB}$ & 1.46(fixed) & 1.46(fixed) & 1.46(fixed) \\ 
${\rm Norm_{CXB}}$ & $4.87 \times 10^{-5}$(fixed) & $4.83 \times 10^{-5}$(fixed) & $1.99 \times 10^{-5}$(fixed) \\
$N({\rm H})_{\rm high} (10^{21}~{\rm cm^{-2}})$ & 30(fixed) & 30(fixed) & 30(fixed) \\ 
$kT_{\rm high} ({\rm keV})$ & 2.5(fixed) & 2.5(fixed) & 2.5(fixed) \\ 
$A_{\rm high}(Z_{\sun})$ & 0.3(fixed) & 0.3(fixed) & 0.3(fixed) \\ 
${\rm EM_{high}}$ & $(6.4\pm2.3) \times 10^{-4}$ & $(8.3\pm2.2) \times 10^{-4}$ & $(2.7^{+0.9}_{-1.0}) \times 10^{-4}$ \\ 
$N({\rm H})_{\rm mid} (10^{21}~{\rm cm^{-2}})$ & $7.09^{+0.72}_{-0.74}$ & $6.19^{+0.70}_{-0.74}$ & $7.48^{+0.79}_{-0.82}$ \\ 
$kT_{\rm mid} ({\rm keV})$ & $0.643^{+0.072}_{-0.074}$ & $0.643^{+0.078}_{-0.041}$ & $0.639^{+0.069}_{-0.058}$ \\ 
$A_{\rm mid}(Z_{\sun})$ & 0.3(fixed) & 0.3(fixed) & 0.3(fixed) \\ 
${\rm EM_{mid}}$ & $(1.40^{+0.35}_{-0.31}) \times 10^{-3}$ & $(1.13^{+0.30}_{-0.26}) \times 10^{-3}$ & $(6.5^{+1.9}_{-1.6}) \times 10^{-4}$ \\ 
$N({\rm H})_{\rm low} (10^{22}~{\rm cm^{-2}})$ & 0(fixed) & 0(fixed) & 0(fixed) \\ 
$kT_{\rm low} ({\rm keV})$ & 0.1(fixed) & 0.1(fixed) & 0.1(fixed) \\ 
$A_{\rm low}(Z_{\sun})$ & 1.0(fixed) & 1.0(fixed) & 1.0(fixed) \\ 
${\rm EM_{low}}$ & $(2.56\pm0.39) \times 10^{-4}$ & $(2.43\pm0.38) \times 10^{-4}$ & $(8.7\pm2.7) \times 10^{-5}$ \\ 
$E_{1}({\rm keV})$ & 1.49(fixed) & 1.49(fixed) & 1.49(fixed) \\
${\rm Norm_{1}}$ & $(4.08\pm0.29) \times 10^{-5}$ & $(3.83\pm0.27) \times 10^{-5}$ & $(1.63\pm0.17) \times 10^{-5}$ \\
$E_{2}({\rm keV})$ & 1.75(fixed) & 1.75(fixed) & 1.75(fixed) \\
${\rm Norm_{2}}$ & $(6.3^{+2.3}_{-2.2}) \times 10^{-6}$ & $(1.25\pm0.23) \times 10^{-5}$ & $(4.5\pm1.4) \times 10^{-6}$ \\
$E_{3}({\rm keV})$ & 0.65(fixed) & 0.65(fixed) & 0.65(fixed) \\
${\rm Norm_{3}}$ & $(1.80\pm0.43) \times 10^{-5}$ & $(1.68\pm0.41) \times 10^{-5}$ & $(7.0^{+1.7}_{-1.6}) \times 10^{-6}$ \\
$\Gamma_{\rm SP,bg}$ & 0.21(fixed) & 0.21(fixed) & 0.21(fixed) \\
${\rm Norm_{SP,bg}}$ & $3.33 \times 10^{-4}$(fixed) & $3.33 \times 10^{-4}$(fixed) & $3.33 \times 10^{-4}$(fixed) \\
$\Gamma_{\rm SP,src}$ & 0.21(fixed) & 0.21(fixed) & 0.21(fixed) \\
${\rm Norm_{SP,src}}$ & $1.11 \times 10^{-3}$(fixed) & $1.07 \times 10^{-3}$(fixed) & $4.51\times 10^{-3}$(fixed) \\ 
$\chi^{2}$/DOF & 271.8/225 & 304.4/244 & 138.8/116 \\
\hline
\end{tabular}
\tablenotetext{}{{\bf Notes:} 
$N({\rm H})_{\rm high}$/$kT_{\rm high}$/$A_{\rm high}$/${\rm EM_{high}}$,
$N({\rm H})_{\rm mid}$/$kT_{\rm mid}$/$A_{\rm mid}$/${\rm EM_{mid}}$, and
$N({\rm H})_{\rm low}$/$kT_{\rm low}$/$A_{\rm low}$/${\rm EM_{low}}$,
are absorption/temperature/abundance/emission measure of the
high-, middle-, and low-temperature plasma models for GRXE (and local diffuse X-ray emission), respectively. 
The emission measures are given as the value
integrated over the line of sight and the region,
$\frac{\Omega}{4\pi} \int n_{\rm e} n_{\rm H} ds$ (where $n_{\rm e}$ and $n_{\rm H}$ are the electron
and hydrogen density, respectively, and $\Omega$ is the solid angle) in the unit of ${\rm 10^{14}~cm^{-5}}$.
The $N({\rm H})_{\rm CXB}$, $\Gamma_{\rm CXB}$, and ${\rm Norm_{CXB}}$ are the 
absorption, photon index of the power-law model, and intensity (${\rm photons~s^{-1}~cm^{-2}~keV^{-1}}$ at 1~keV)
integrated over the region for the CXB model, respectively. 
$E$ and ${\rm Norm}$ with subscripts 1, 2, and 3 are line center energy and the intensity 
(${\rm photons~s^{-1}~cm^{-2}}$) of gaussian, respectively,
to model the fluorescent background lines and the SWCX.
$\Gamma_{\rm SP,bg}$/${\rm Norm_{SP,bg}}$ and
$\Gamma_{\rm SP,src}$/${\rm Norm_{SP,src}}$ are the power-law index and the normalization of the
residual soft-proton contamination in the background region and the source region, respectively.
They are convolved with the response matrix of diagonal unity elements.
Errors are calculated for single-parameter 90\% confidence limit.
See also Table~3 for the spectral parameters of the PWN emission.
}
\end{center}
\end{table}

\clearpage

\listofchanges

\end{document}